**Synthesis of One Atom Thin, Two-Dimensional Gold Films and Their Novel Properties**

Sudhir Kumar Sharma,[1] Renu Pasricha,[2] James Weston,[2] Florian Stumpf [3], Thomas Blanton[4], and Ramesh Jagannathan[1*]

[1] Engineering Division, New York University Abu Dhabi, Abu Dhabi, United Arab Emirates.

[2] Core Technology Platform, New York University Abu Dhabi, Abu Dhabi, United Arab Emirates.

[3] Park Systems Europe Gmbh, Janderstrasse 5, DE-68199 Mannheim, Germany

[4] International Centre for Diffraction Data, 12 Campus Blvd., Newtown Square, PA19073, United States of America

**\*Correspondence address: Ramesh Jagannathan** (Email: rj31@nyu.edu )

**Abstract:**

Though significant advances have been made in the field of metal nanostructures, researchers are yet to synthesize one atom thin two-dimensional (2D) gold nanostructures. We report, for the first time, a technique to synthesize one atom thin gold films and *membrane-like* porous films on silicon and sapphire. These films were essentially liquid at room temperature. Current-Voltage (IV) spectroscopy using atomic force microscopy (AFM) revealed a typical Schottky behavior with a very high turn on (knee) voltage at +4.15V indicating that these 2D gold structures are semi-conductors. Nanorings comprising self-assembled one atom thin gold structures on sapphire were also observed. Subjecting **one** nanoring in a group of several nanorings to a point force of 2.25 µN for 120 seconds resulted in the creation of a *mirror* structure, accompanied by every nanoring in the group to simultaneously create its individual *mirror* structure(s). These *mirror* structures had an apparent spatial and temporal correlation to each other. They had a finite life time of a few hours and the process was repeatable to a great degree of precision.

**Introduction:**

Two dimensional (2D) materials, which are single crystalline layer of atoms or molecules, are of interest to researchers because of their applications across varied disciplines. [1–9] It is difficult to overstate their impact on both basic and applied science. Advances in 2D materials have led to a burst of research activities in the so-called field of vdW crystals and spintronics with an emphasis on the potential discovery of a room temperature vdW ferromagnetic spin source.[10–15] More recently, Wang et al have exploited the so-called Berry curvature memory to electrically induce stable stacking shifts in a three-layer tungsten ditelluride. [16] Recent advances in van der Waals (vdW) heterostructures comprising a vertical stack of different 2D materials have infinitely expanded the scope of custom designing a variety of devices with novel properties by manipulating the material type and stacking sequence.[17,18] More recently, researchers have discovered that interaction between two atomically thin layers of tungsten diselenide/tungsten disulfide qualitatively changed their electronic band structure. [19]

The obvious importance of 2D materials focused our interest on 2D metal nanostructures. There have been significant advances in the general field of metal nanostructures, [8,9,20,21] especially gold, due to its potential high impact applications in areas such as photonics, electronics, optical-sensing/imaging, and drug-delivery.[22,23] Researchers have so far developed a range of gold nanostructures using various biological, photochemical, and wet chemical methods.[6,24–28] Recently, Ye et al, for example, reported a facile strategy to synthesize free-standing two atomic layer thick gold structures (AuNSW)[29] with enhanced catalytic properties. However, no one has so far succeeded in synthesizing a two-dimensional crystalline nanostructure of gold, consisting of a single layer of atoms.

In our research, we specifically set out to close this gap in the field. Our experimental approach was based on the well-known phenomenon of "melting point depression" in high surface to volume ratio materials. The Lindemann criterion dictates that a solid should start to melt when the amplitude of its atomic vibrations and nearest-neighbors distance ($R_{IN}$) become comparable[30].

In other words, a solid will melt if the ratio,

$$\gamma = \frac{(u^2)^{1/2}}{R_{1N}} \quad \text{where} \quad (u^2) = \frac{4h^2 T}{M k_B \Theta_s^2}$$

reaches an optimum value $\gamma_m$. In the above equation, the term ($u^2$) is the mean square of the displacement of an atom from its equilibrium position. h, is the Planck's constant, M, is the atomic mass of the metal, $k_B$, is the Boltzmann's constant and $Q_s$, is the surface Debye temperature. [31] Killean et al reported a linear decrease in the melting point of metals with a decrease in their Debye temperatures.[32] It is well established that the vibrational amplitude of surface atoms of a crystalline material could be up to 100% higher than their bulk value, resulting in up to a 50% decrease in their surface Debye temperatures compared to their bulk temperatures. In other words, crystal surfaces would start to melt at temperatures that are significantly lower than their standard bulk melting temperatures. Frenken et al were the first to report a partial disordering (melting) of the Pb (110) surface starting at 75% below its bulk melting point.[33] These findings were corroborated by Breuer et al.[34] Ma et al reported a range of surface Debye temperatures for polycrystalline gold films from 83K to 121K as compared to their bulk value of 165K.[35] Santos et al observed solid state de-wetting of gold films resulting in the formation of crystallites and islands growth below 360°C.[36] At temperatures greater than 360°C, they observed the gold film to melt and diffuse on the surface to the crystallites.

Based on the above observations, we hypothesized that exposure of a gold thin film sample on a substrate (e.g. sapphire, silicon) to an optimum temperature that results in the (partial) melting /disordering of its surface, for a short duration of time (for example, t = 2 - 60 seconds), is likely to result in the following phenomenon: the gold film in contact with the substrate would start to melt first and recede from the substrate-film interface; the resulting decrease in surface energy and the associated higher melt/solid equilibrium temperature, would lead to re-crystallization of gold on the substrate. This process of recurrent melting and re-crystallization is likely to result in instability at the film/substrate interface. This instability would be resolved if the re-crystallization process resulted in a 2D structure of gold (which, by definition, does not have any "bulk").

Our experiments based on this hypothesis successfully led to the observation of the following *one atom thin* crystalline gold nanostructures: continuous films and membrane-like porous films, 2D honeycomb type superlattices, and arrays of high aspect ratio (~ 2500) hexagonal disc. These structures were essentially liquid at room temperature. We also observed that the gold 2D structures on sapphire self-assembled to form nanorings that exhibited several unique and difficult-to-explain phenomena. For example, subjecting **one** nanoring in a group of nanorings to a point force of 2.25 μN for 120 seconds resulted in it nucleating a *mirror* structure. This was simultaneously accompanied by every nanoring in the group to spontaneously nucleate its individual *mirror* structure(s) with an apparent spatial/temporal correlation to each other. The *mirror* structures had a finite life time of a few hours and the process was repeatable to a great degree of precision. Electrical characterization of the nanostructures showed that they are semiconductors with a high turn on voltage of +4.15V.

**Results:**

As a first step, we determined the optimum temperature through a series of screening experiments in which a thin film of gold (e.g. X-ray Reflectivity (XRR) thickness ~ 66 nm (Figure S1) on a single crystal sapphire substrate was exposed to temperatures of 350°C, 425°C and 475°C for 5 seconds. AFM images of the heat-treated films are shown in Figure S2. We observed typical solid-state de-wetting of the films at 350°C and 425°C when the film separated into several small crystallites. No crystallites were formed at 475°C and nanoring structures were observed instead. AFM images of the nanoring structures formed when the gold film was heat treated to 475°C for 2s, 5s, 30s and 60s, respectively, are shown in Figure S3. Detailed characterizations of the films revealed that the sapphire and silicon substrates were primarily covered with one atom thin continuous films and membrane-*like* porous films. In this manuscript, we report our experimental results for gold thin films deposited on sapphire and silicon substrates, respectively and heat treated to 475°C.

In Figure S4, we show the X-ray diffraction (XRD) of a gold film on sapphire heat treated to 475°C for 60s, confirming the presence of gold with (200) and (220) diffraction peaks. Deposition of gold films typically results in a strong (111) diffraction peak (~38.5 degrees 2θ using Cu Kα radiation). Absence of the gold (111) diffraction peak in these samples would imply a significant change in the morphology of the gold due to the 475°C for 60s thermal processing. Sample characterization using secondary electron microscopy was hampered by beam sensitivity (Figure S5). Since, sapphire and silicon substrates were used in our experiments, TEM was ruled out as a primary method of study. However, TEM was effectively used for chemical and structural characterizations. While AFM was the primary method of characterization of the gold structures reported in this study, *key results were corroborated by electron microscopy.* The Park Atomic Force Microscope (Park NX10) instrument used in our experiments was capable of operating in a "true" non-contact mode to facilitate non-invasive image capture. The experimental results reported in this manuscript are for samples which were treated to 475°C.

A. **One atom thin, 2-dimensional gold films on silicon and sapphire substrates:**

   a) **TEM Analysis:**

In Figure 1 we show the results from the TEM characterization of the heat-treated gold sample on sapphire. A facile process was implemented for the direct transfer of the gold nanostructures/film from the sapphire substrate to the selected target for HRTEM imaging, the Quantifoil with 1.2μm holey carbon film coated Cu grids. This direct route of transferring the film ensured contamination free transfer. The details of the process are described in the methods section. Figure 1a is a low magnification STEM image of the gold sample and we can clearly see the thin gold films across the entire imaging field. There are indications of some of the thin gold films rolling up into needle/tube like forms during the transfer process marked by white arrows. The uneven and non-uniform shapes are similar to that reported by Ye et al who had synthesized two-atom thick gold films[29]. 1b shows the high magnification STEM image of a region in 1a marked by the white square labelled b. The bright area in the middle is a thick stack of 2D gold films. The surrounding large grey area shows a highly ordered, 2D gold superlattice structure with well-defined edges (white arrow) and morphology. The black arrows point to rolled up 2D gold films. The white rectangle highlights a well-defined 2-D gold film that is rectangular in shape. 1c and 1d are the high magnification STEM images of a region in 1a marked by the white squares labelled c and d, respectively. They confirm the quality/integrity and extent of the 2-D gold films with well-defined edges produced by our process.

Figure 1e shows the selected area electron diffraction (SAED) pattern of a gold thin film shown in 1a. The pattern has a six-fold symmetry indicating the {111} zone axis. Diffraction associated with the fractional lattice planes 1/3{422}, occur in addition to the expected (111), (200) and (220) reflections of fcc gold, on the basis of their respective d-spacing of 2.40 Å, 2.05 Å and 1.467 Å (JCPDS card No. 04-0784). The observed d-spacing of 2.8 Å corresponds to forbidden 1/3{422} reflection. The colored circles correspond to the reflections from the various lattice planes. According to Cherns, the 1/3{422} diffraction spot could only be observed under either of the following two conditions - an incomplete FCC lattice due to the presence of only one atomic layer or because of twinning in the film with less than three atomic layers[37]. We do not see any evidence of twinning. In combination with this result, we will unequivocally confirm the presence of one atom thin gold films by AFM characterization (Figure 2).

High resolution surface imaging in bright field mode was used to conduct the atomic position analysis for the single monolayer of Au (1 1 1) surface. The incident beam direction was normal to the specimen surface, and bright field imaging was used for the atomic resolution imaging. In Figure 1f, we show a HRTEM image of an area of Au film in (111) zone axis. The fringe spacing of the image within the white circle was measured to be ~2.4 Å consistent with the fcc gold structure. Thermal expansion of gold crystals should be taken into consideration when discussing the lattice constant of the structures subjected to elevated temperatures. The lattice spacing was measured using the IFFT (inverse fast Fourier transformation) image (Figure 1f inset) reconstructed from the bright field image in Figure 1f.

In Figure 2 we show the results for gold films on sapphire (a – c and i-k) and silicon (d - h) substrates, respectively. In Figure 2a we show an AFM image of stacks of gold films on sapphire with distinct triangular morphologies with well-defined but ragged, *fractal like* edges, after heat treatment at 475°C for 60 s. Figures 2b and 2c are higher magnification images of two selected regions in Figure 2a. The images confirm the presence of very thin films with well-defined edges and morphologies. Height measurements along the white lines show 1.5 Å – 2.0 Å thinness, confirming, in unequivocal terms, that these are one atom thin, 2-dimensional gold films. These results confirm that the forbidden 1/3{422} reflection observed in Figure 1b was due to one atom thin gold film.

In Figure 2d, we show an SEM image of a stack of very thin gold films on silicon that were heat treated to 475°C for 10 s. In Figure 2e, we show an AFM image of stacks of thin gold films on silicon with clearly defined shapes and edges that are qualitatively consistent with the SEM observations. The white dotted line highlights a hexagonal shaped film. In Figure 2f, we show a higher magnification AFM image of a region in Figure 2e. Once again, the image reveals very thin gold films with well-defined morphologies and the white dotted lines highlight a hexagonal shaped film. The height profile along the white line 1 measured the thinness of a film to be 1.5 Å – 2.0 Å, confirming that these are one atom thin 2D gold films. In Figure 2g, we show an AFM image of a region of the gold film on silicon. It clearly shows a highly ordered, 2D array of one atom thin gold structures as confirmed by the height profile (~2.0 Å). The 2D array structure is confirmed by the SEM image shown in Figure 2h. In Figure S6A we show a larger (3μmx3μm) area of the silicon substrate covered with gold membrane film. A higher magnification AFM image of a select region in Figure S6A is shown in Figure S6B. Honeycomb type 2-dimensional gold structures were also observed on 475 °C heat treated films on sapphire (Figure 2i). Height profile measurements of a high magnification AFM image (Figure 2j) of a region in Figure 2i revealed these to be one atom thin gold structures (~2.0 Å). We also observed highly ordered 2D hexagonal gold structures on sapphire as shown in Figure 2k. Height profile measurement across one such hexagonal structure unequivocally confirmed that it is one atom thin (~2.0 Å). In Figure S6C and S6D, we show two other larger areas (2μmx2μm) of the sapphire substrate covered with hexagonal two-dimensional gold structures. In Figure S7, we show the EDS analysis of the two-dimensional film on silicon, confirming that they are indeed gold.

Park AFM system was used to obtain the IV (Current-Voltage) spectrograph for the two-dimensional gold film on silicon. We observed a typical Schottky behavior (Figure 2l) for semi-conductors with a very high turn-on (knee) voltage of +4.15 volts. This is much higher than that for silicon (0.7 V) and GaN (1.6), a much-studied wide band gap (WBG) material.[13–15] In contrast, the I/V spectroscopy of the corresponding (as deposited) gold film showed a typical Ohmic behavior that is characteristic of metals (Figure S8).

In Figure S9a, we show a representative AFM image of nanoring structures formed by self-assembly of some of the 2-dimensional gold films. The detailed description of nanoring formation is included in the next section. AFM image of one of the regions between two nanoring structures marked by the white circle is shown in Figure S9b, revealing the presence of a stack of one atom thin gold films with a height/thinness of ~2.0 Å. In Figure S10a we show a higher magnification

AFM image of a nanoring and the region around its base indicating the presence of a number of thin gold films. We measured the height /thinness of one of the films (Figure S10b) to be 1.3 Å indicating that it is a one atom thin gold structure.

## B. Nanoring structures:

### a. *Characterization of Nanorings*

In our experiments, we observed circular open top dome-like growth structures of gold films (Figure S11). These growth structures were found to be rising from their edges and pinned to the substrate, as shown in Figure S11a and S12. Over a period of time, these structures appeared to coalesce with each other and grow stacks of larger circular arcs, eventually leading to well-formed asymmetric nanorings (Figure S11b-g, Figure S13). High-resolution AFM image of a selected area of the nanoring (Figure S11g – marked with a white arrow) is shown in Figure S11h. It is evident that the nanoring is a composite assembly comprising several smaller discs/lamellae, as their basic building blocks. The height profile measured across these lamellae are found to be approximately 1.7Å implying one atomic layer of gold (shown in Figure S11i).[8,27,29,38,39]

Our calculations based on the AFM images indicate that approximately 10-20% of the deposited gold was converted to nanorings upon exposure to 475°C for 60 seconds, the rest remaining on the substrate as one atom thin porous and non-porous films. The observation of the various stages of growth of the nanorings is consistent with the thermodynamic framework of partial melting/re-crystallization of thin films discussed earlier.

AFM images of a typical nanoring (Figure S11g, S14a) clearly showed that each ring is actually composed of several lamellae (Figure S11h and S14b). The base of the nanorings on the substrate was also found to be covered with similar, lamellae structures. STEM imaging confirmed the AFM results (Figure S14c) and that the nanoring is indeed gold (Figure S14d, g and h). High resolution TEM (HREM) (Figure S14e and f) confirmed a gold lattice spacing of 2.8 Å. [29,40–42] In Figure S15, we show elemental X-ray spectral mapping (EDS) of gold across individual nanoring structures for several nanorings confirming that these are indeed gold structures.

### b. Unusual nanoring phenomena:

#### a. Spatial Memory

During the course of our characterization of the two-dimensional gold nanostructures, we also observed a number of interesting phenomena that we found difficult to explain. At room temperature, we subjected one of the nanorings, to a momentary force (100 nN), several times in short succession, using the Force/Displacement (F/D spectroscopy) technique in the AFM (Figure 3) [43–45]. The force was applied at selected locations, shown by the red (+) marks on the image in Figure 3a. We show the AFM scan collected for the same sample field before (Figure 3a) and after the application of the force (Figure 3b-d), at various time intervals. The application of force to one nanoring in a group of nanorings (referred to as *mother- rings*) resulted in a significant increase in volume of each *mother-ring* in the group and formation of several so-called *daughter-rings* associated with each *mother-ring* (Figure 3b).

Over a period of time, without the application of any additional external force or stimulus, each *daughter-ring* returned to its own respective *mother-ring* and re-assembled in its original form, orientation and size (Figure 3c and 3d). The final AFM image of all the *mother-rings in the*

*group* (Figure 3d) were identical to that observed prior to the application of the force (Figure 3a). Each *daughter-ring* only returned to its respective *mother-ring* during the re-assembly process, even if two *daughter-rings* from two different *mother-rings* were in significant physical contact with each other (Figure S16). In our experiments, we did not encounter a situation where some of the *daughter-rings* had completed the re-assembly process, while others have not. In Figure 3e and 3f, we show backscattered electron images of *mother-rings* and their *daughter-ring* replicas, corroborating the AFM results on the formation of *daughter-ring* structures (shown in Figure 3b). This series of events is puzzling.

The series of events that we described above (in Figure 3) was reproduced in the same region as well as other regions of the same and other samples with different AFM tips/cantilevers. For example, the effect of force on the nanoring shown in Figure 3a was simultaneously and uniformly experienced by all the nanorings within the 100 μm x 100 μm scanner bed. In Figure S17, we show that nanorings located 45μm away from the point of application of the force (Figure 3a) exhibited similar phenomenon as shown in Figure 3. In each experiment, we had imaged several nanorings across the entire scanner area and found all of them to exhibit the same phenomenon.

Having confirmed the series of events as described, we next sought to rule out any artificial or instrument-based artifacts as explanations for the observed phenomena. In addition to the corroboration by backscattered electrons microscopy, the time dependent appearance and disappearance of the *daughter-ring* fragments and the repeatability of the phenomenon over time on the same sample as well as on several other samples **ruled out** any AFM imaging related artifacts. Additionally, we also eliminated any residual concern regarding AFM imaging artifact by scanning the same sample region at two different angles, namely, 0 and 90 degrees. The rotation of the image features for the 90-degree scan (shown in Figure S18) clearly confirmed the absence of any imaging artifact.

c. *Stress response behavior: Mirror Nanostructures*

Based on the above unique observations, we studied the room temperature, force response behavior of the nanorings in detail and discovered their unique tendency to reproducibly create *mirror* structures that have never been previously observed or associated with inanimate systems. We applied a one-time force of 2.25 µN to one nanoring in a group of several nanorings for 120 seconds, in the location marked by the white (+) mark in Figure 4a. This led that nanoring to nucleate a *mirror* structure. This was simultaneously accompanied by every nanoring in the group to nucleate its individual *mirror* structure, as shown by the white dotted arrow lines in Figure 4b. Subsequent imaging showed a substantial growth of all the *mirror* structures (Figure 4c). All the *mirror* structures were oriented at a 30 degrees angle to the horizontal plane as marked in Figure 4c and were located at a distance of 0.64 µm from their respective *mother* nanoring. Over a period of a few hours, the *mirror* structures disappeared (Figure 4d). This remarkable stress response mirroring behavior, (i.e. creation of *mirror* structures) was predictably reproduced as shown in Figure 5. First application of a 2.25 µN force for 120 seconds to one nanoring at the white (+) mark in Figure 5a resulted in the spontaneous creation of *mirror* structures (Figure 5b) by all the nanorings in the group, similar to that observed in Figure 4c. Over a short time period, the *mirror* structures disappeared. Subsequent second application of the 2.25 µN force for 120 seconds to the same nanoring at the same white (+) mark location (Figure 5c), resulted in the spontaneous generation of *mirror* structures (Figure 5d) that were identical to those observed in Figure 5b. Again, over a period of an hour, the *mirror* structures disappeared. Subsequent third application of the force to the same nanoring at the same location (Figure 5e) again produced the same *mirror* structures (Figure 5f). Identical stress response mirroring behavior was demonstrated at another sample location (Figure S19).

We extended our study to examine the stress response *mirroring* behavior of the *mirror* structures themselves. In Figure 6a we show the AFM image of a region from Figure S20. Figure 6b is a higher magnification image of the region within the white square in Figure 6a. We applied a force of 2.25μN for a period of 120 seconds each, at the three white (+) marked locations on the middle *mirror* structure in Figure 6b. This resulted in splitting of the middle *mirror* structure into two equal parts as shown in Figure 6c. In Figure 6d, we show a higher magnification image of the region within the yellow square in Figure 6a imaged after the application of the force described in Figure 6b. It is remarkable to observe that the middle *mirror* structure in Figure 6d has split into two equal parts as well. We observed the same phenomenon of splitting in the middle *mirror* structure within the red square in Figure 6a as well.

In Figure 7 we show further evidence of the remarkable phenomenon of *mirroring* behavior in the *mirror* structures. In 7a, we show the AFM image of nanorings and their *mirror* structures. The *mirror* structures are identified by the *tail* at their bottom right side. We chose two *mirror* structures, within the yellow and white squares, respectively, for our study. Higher magnification images of these structures are shown in 7b and 7c. We applied a force of 2.25 μN for 120 seconds at the *tail* location at the white (+) mark shown in 7b. 7d, 7e and 7f are AFM images of the regions of interest after the application of the force. The stress response behavior of the *mirror* structure, namely, the disappearance of the *tail* due to the applied force, was simultaneously *mirrored* by all the *mirror* structures in the imaged field.

An AFM image of the clean sapphire substrate, for reference purposes, is shown in Figure S20.

### d. *Liquid like gold films at room temperature*

In Figure 8 we demonstrate the *liquid-like* characteristic of these 2-dimensional gold films at room temperature. In Figure 8a we show an AFM image of nanorings and a long strip of thick gold film in the middle (yellow rectangle/arrows) and a smaller film to the left (white rectangle/arrows). The presence of cascading liquid wave fronts on the film in the middle and on the nanoring around the white (+) mark, indicated by the white arrows are clear and distinct. We applied a point force of 2.25 µN for 120 seconds to the nanoring surface at the white (+) mark in Figure 8a. We imaged the region immediately afterwards (Figure 8b) and found that several of the liquid wave fronts (within the yellow rectangle) have coalesced into one distinct wave front that had moved forward in a significant manner. The liquid film in the region inside the white rectangle appears to have de-wetted from the surface and receded, indicating surface tension effects. The arrows (yellow/white) indicate the wave fronts. We again applied the same force of 2.25 µN for 120 seconds at the white (+) mark resulting in the generation of new liquid wave fronts (yellow rectangle) as shown in Figure 8c. We observed that the liquid in the region marked by white rectangle (Figure 8c) has spread out. We repeated the application of the force for the third time which led to the spreading of the liquid film in the middle of the image (yellow rectangle) completely covering the region (Figure 8d). The liquid film inside the white rectangle had de-wetted again and receded back.

**Discussion**

We have successfully prepared two-dimensional gold nanostructures consisting of a single layer of atoms on silicon and sapphire. We observed stacks of one atom thin 2-dimensional gold films with well-defined morphologies and ragged *fractal-type* edges and ordered honeycomb type superlattices on both sapphire and silicon substrates. We also observed one atom thin ordered arrays of hexagonal shaped gold structures on sapphire. Transmission electron microscopy (TEM), selective area electron diffraction (SAED) and AFM imaging were used to unequivocally confirm the presence of one atom thin (2D) gold films. Observation of the forbidden reflections from fractional lattice planes 1/3{422} in the SAED from the 2D gold thin films unequivocally confirmed that they are one atom thin structures. 2D ordered gold superlattice structures were observed by all three microscopy techniques, namely TEM, SEM and AFM. The 2D gold films were essentially liquid at room temperature. Some of these 2D gold films self-assembled to form nanorings with novel properties.

While it is well-known that all of gold's low-index surfaces tend to reconstruct resulting in higher surface density and close packed triangular surface structures, the "glue" model force field simulations by Ercolessi et al [46] were the first to explain the *liquid like* characteristics of gold surfaces. For example, explanation of the propensity for gold surfaces to have a large coordination in spite of their large cohesive energy as a relativistic effect due to the lowering of 6s levels relative to 5spd, instead of a structural order. Our experimental findings are the first clear demonstration of the implications and extensions of such predictions for a 2D gold structure.

Electrical characterizations revealed that the gold nanostructures are semi-conductors demonstrating a typical Schottky behavior but with a high turn on (i.e. knee) voltage of +4.15 V. The observation that these structures are not metallic is qualitatively consistent with the expectation that a one atom thin metal lamella (i.e. gold) would have its electrons truly localized/pinned at the atomic level. [13–15] This observation is still surprising from the viewpoint of gold's standard electronic structure because it would imply(?) a full band filling, and a large bandgap, with its 6s electrons. Whether these 2D structures fall under the classification of Mott insulators remain to be explored. It has been suggested that what we are observing might be a Coulomb blockade phenomenon.

Liu et al based on their energy and dynamic analysis calculations concluded that a one atom thin two-dimensional gold with a honeycomb (HG) structure is stable from a thermodynamic and lattice dynamic viewpoint.[47] They showed that the HG structured gold exhibits a covalent bonding characteristic and is a semiconductor with a band gap of 0.1 – 0.3 eV. J.R. Ahn et al studied the electronic structures of the Au-induced one dimensional chain structures on Si(557).[7] Using Angle Resolved Photoemission (ARP) spectroscopy and Scanning Tunneling Microscopy (STM), they characterized the two proximal 1D bands near the Fermi level, one as metallic subject to metal-insulator transition upon cooling and the adjacent one with a band gap at room temperature. These studies are qualitatively consistent with our observation of honeycomb type gold nanostructures and that they are semiconductors.

In summary, one could conclude from the reported results that, in each lamella, there is little or no electron de-localization from the nucleus and the metallic bond (electrostatic) interactions are confined to the one atom thin crystal lattice plane with significant surface anisotropy. This would imply that the interlamellar interaction between the numerous lamellae in the nanorings are likely to be governed by van der Waals forces. In this scenario, each lamella would be a vdW crystal and the nanoring would be a vdW homostructure. This would explain the observed experimental phenomenon of force response of the nanorings, namely increase in volume and uniform release of the *daughter* structures as reported in Figure 2. What is still unresolved is the return of each *daughter* to its respective *mother*, even if two daughters from two different mothers were in significant physical contact with each other and their original spatial orientation memory. The apparent synchronization of the return process among all the *daughters* remains unexplained as well.

The *mirroring* phenomenon as a stress response behavior is extremely novel and puzzling. While this behavior is well known for certain biological systems [48], it is unheard of for inanimate systems. It is important to note that we observed two specific *mirroring* behaviors. First, the *mirroring* behavior of a group nanorings when one of them is subjected to a point force. The second is the *mirroring* behavior of a group of *mirror* structures themselves, when one of them was subjected a point force. At this point of time, we do not claim to have an explanation for this extraordinary behavior, but it is logically indisputable that the phenomenon would require simultaneous and precise transmission of the information related to the highly localized applied point force to a specific structure, to all the *other structures*. What is not obvious is the source of the force/stress experienced by each of these *other structures* triggering their *mirroring* behavior. The question is, *how* did each of these nanostructures in a group precisely and simultaneously experience the applied stress to one *specific nanostructure* in the group and why did they respond by mirroring the behavior of that *specific nanostructure*? It is evident that extensive theoretical and experimental work is needed to establish these exciting new findings on a firm scientific foundation.

**Conclusions:**

We report, for the first time, a technique to synthesize one atom thin gold films and *membrane-like* porous films on silicon and sapphire substrates. We also report the observation of two-dimensional arrays of honeycomb type one atom thin gold superlattices and similar arrays of high aspect ratio (~2500) gold hexagonal lamellae on sapphire substrates. The observations were corroborated by electron microscopy (TEM, SEM), Atomic Force Microcopy (AFM) and selective area electron diffraction (SAED). These nanostructures were essentially liquid at room temperature.

Electrical characterization of these nanostructures revealed that they are semi-conductors with a very high turn on voltage indicating a wide bandgap (WBG) semi-conductor material. We speculate that the nanoring structures comprising self-assembled one atom thin gold lamellae suffer no electron de-localization and their metallic bond (electrostatic) interactions are confined to the crystal lattice plane with significant surface anisotropy. It has been suggested in the literature that they might exhibit covalent bond characteristic[47]. This would imply that each lamella is a vdW crystal and a collection of them self-assemble to create vdW homostructure, namely a nanoring (Figure S11h).

The highly repeatable and precise response of the nanorings to an applied local force imply two important characteristics, namely, an apparent *homing action* and *mirroring behavior*. These remarkable characteristics are unprecedented in non-biological systems and have the potential to fundamentally disrupt our understanding of the physics governing these behaviors. Some future applications would include the concept of *robots on chip* and ultra-sensitive detection of seismic activity. We expect these phenomenological observations to stimulate significant fundamental research.

**Materials and Methods**

   *A. Materials:*

Precursor material, namely, gold wires were purchased from Sigma-Aldrich (99.99%, highest available purity grade). Highly polished silicon (76.2 mm in diameter, N-type, SSP Prime <100>) and sapphire substrates (0.33 mm thickness, Single crystal, DSP Prime C-plane <0001> off M-plane <1-100> 0.2 ± 0.1°), were purchased from University Wafer Inc. USA.

   B. *Deposition of Gold Films*:

Thermal evaporation process was used to deposited the gold thin films from a high purity gold wire onto the silicon and sapphire substrates. First, these substrates were ultrasonically cleaned with isopropyl alcohol (IPA) and dried with a nitrogen-spraying gun. The substrates were then loaded into a plasma cleaning system (PDC-002, Expanded Plasma Cleaner, Harrick Plasma, USA) for a duration of 5 minutes. Freshly cleaned substrates were transferred to the thermal evaporation chamber of Denton Vacuum LLC 502 B system. The base pressure of main chamber was ~ $3 \times 10^{-7}$ torr. Prior to the evaporation process, a lower current of ~50A was applied for 3 minutes for preconditioning the tungsten boat. Film deposition rate of 1Å/sec at 72 Å was calibrated by an in-built quartz crystal monitor (QCM) and maintained constant throughout the process. Gold coatings with different (QCM, Quartz Crystal Monitor) thicknesses (e.g. 10nm, 25nm, 30nm, 50nm (66nm by X-ray reflectivity), 100nm, 200nm) on single crystal sapphire substrate were deposited. Subsequently, the deposited films diced into 0.5 mm x 0.5 mm square shaped samples and post-deposition heat treated in a preheated oven (Carbolite 1200, UK). Post-deposition heat-treatment temperature ranged from 350°C – 475°C in air. The duration of heat treatment were 2s, 5s, 30s, and 60s, respectively.

C. *AFM characterizations of Gold Films:*

### 1) Surface Topography

Atomic force microscope (Park NX10, Park Systems, Korea) was used in true non-contact mode for morphological characterization. This unique scan mode prevents potentially invasive tip-sample interaction during a scan. AFM scans were collected using Park SmartScan™ software (Park Systems, Korea) using ultra-sensitive (Super Sharp Standard NCH cantilevers, High Density Carbon tip, Manufacturer: Nano World AG, Switzerland) with a typical tip height of 10-15μm and 2nm tip radius, under ambient conditions. The physical dimensions of these cantilevers were 125μm (length), 30μm (width) and 4μm (thickness), respectively. The force constant value and resonant frequency were 42Nm$^{-1}$ and 320 kHz, respectively.

### 2) F-D spectroscopy in Park AFMs

In addition to conventional AFM imaging, Park AFM has in built capabilities to precisely measure the nano-scale mechanical properties of materials, namely, nano-indentation technique (load-displacement (P-H) curve) and an additional capability referred to as the Force vs Distance (F/D) Spectroscopy. F/D spectroscopy measures the vertical force applied by the tip to the surface during contact-AFM imaging, through the deflection of the cantilever as a function of the extension of the piezoelectric scanner. Typically, this technique employs a significantly smaller force than that used in conventional AFM nano-indentation experiments. In F/D spectroscopy, the dependence of the cantilever deflection on the extension of the piezoelectric scanner is directly correlated to the tip-sample interaction forces reflecting a surface mechanical property. In principle, the technique could be used to measure local variations in the elastic properties of the surface.

### 3) I-V spectroscopy in Park AFM

We used Park NX10 AFM equipment to carry out IV (Current vs. Voltage) spectroscopy on distinct three-dimensional gold structures. The low noise design feature of the system's conductive AFM (C-AFM) option enabled us to detect extremely small changes in a sample's electronic characteristics with great precision. The internal amplifier was run at the $10^6$ V/A Gain. A CDT-NCHR (Conductive Diamond Coated Tip - Non-Contact/Tapping Mode - High Resonance Frequency - Reflex Coating) cantilever was used for these experiments. IV spectroscopy was carried out in three different locations on the sample, from 0V -> 5V -> -5V -> 0V. The entire sweep took 5 seconds.

### D. *TEM sample preparation and imaging:*

A facile route was implemented for the direct transfer of the gold nano-structures from sapphire substrate to the selected target for HRTEM imaging, the Quantifoil with 1.2μm holey carbon film coated Cu grids. This direct route of transferring the film ensured contamination free transfer. During the direct transfer, the support to the gold film is provided by TEM grid's carbon film. To bond the gold and carbon of the TEM grid, the grid is placed on top of gold and a drop of isopropanol (IPA) is gently placed on top of the grid ensuring that both the grid's carbon film and the underlying gold film is wet. The surface tension draws the gold and carbon together into close contact as IPA evaporated[49]. Residual IPA was evaporated at ambient conditions over a period of 24 hours.

Characterization of transferred gold film on the TEM grids was performed using a Talos F200X FEG Transmission Electron Microscope with a lattice-fringe resolution of 0.14nm at an accelerating voltage of 200 kV equipped with CETA 16M camera. High resolution images of periodic structures were analysed using TIA software.

**Acknowledgments:** We acknowledge Prof. Erio Tossati (SISSA (Scuola Internazionale Superiore di Studi Avanzati) and ICTP (The Abdus Salam International Centre of Theoretical Physics)) for his keen insights and comments. We acknowledge the valuable support provided by the Core Technology Platforms at NYU Abu Dhabi for the use of the instruments. We acknowledge the editorial support provided by Phillip Rodenbough at NYU Abu Dhabi.

**Funding:** Funding for this work was provided by NYU Abu Dhabi.

**Author contributions:** RJ developed the concept behind the experiments, designed the experimental plans and wrote the manuscript. SKS carried out all the experiments and SKS and RJ shared the AFM characterization responsibilities and analysis. RP did the TEM characterization. JW did the XRD characterizations. FS carried out the C-AFM measurements. TB analyzed the XRD data and wrote the sections on XRD.

**Competing interests:** The authors declare no competing interests.

**Data and materials availability:** All data generated or analyzed during this study are included in this published article (and its supplementary information files).

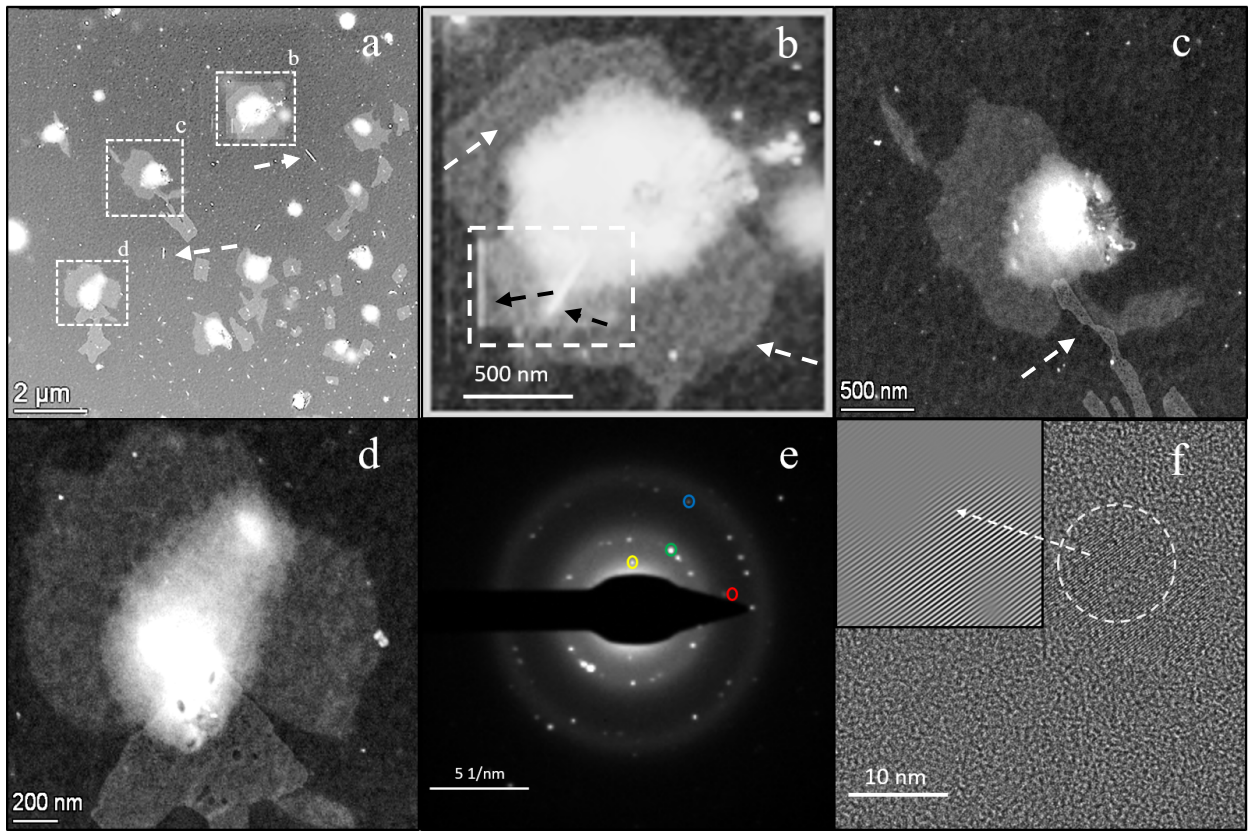

**Figure 1: Confirmation of 2D gold film by TEM high resolution lattice imaging (HREM) and selective area electron diffraction (SAED):** 1a is a low magnification STEM image of the heat-treated gold sample on sapphire that has been transferred to the TEM grid. The light grey areas show the extent of the 2D gold film and the high intensity areas represent localized stacks of 2D gold films. The bright needle/rod like features (white arrows) represent rolled up 2D gold films during the transfer process. 1b shows the high magnification STEM image of a region in 1a marked by the white square labelled b. The bright area in the middle is a thick stack of 2D gold films. The surrounding large grey area shows a highly ordered, 2D gold superlattice structure with well-defined edges (white arrow) and morphology. The black arrows point to rolled up 2-D gold films. The white rectangle highlights a well-defined 2D gold film that is rectangular in shape. 1c and 1d are

the high magnification STEM images of a region in 1a marked by the white squares labelled c and d, respectively. The flexible nature of the films (white arrow) is evident in 1c. 1e shows the SAED from a 2D gold film. The colored circles correspond to the lattice planes, (220)/Blue, (200)/Red (111)/Green and 1/3{422}/Yellow, which is the *forbidden reflection*, respectively. 1f is a high-resolution electron image of a thin gold film showing lattice imaging (white circle). The inset is the inverted fast Fourier transform (IFFT) image of the lattice inside the white circle. It shows a periodic d-spacing of 2.8 Å.

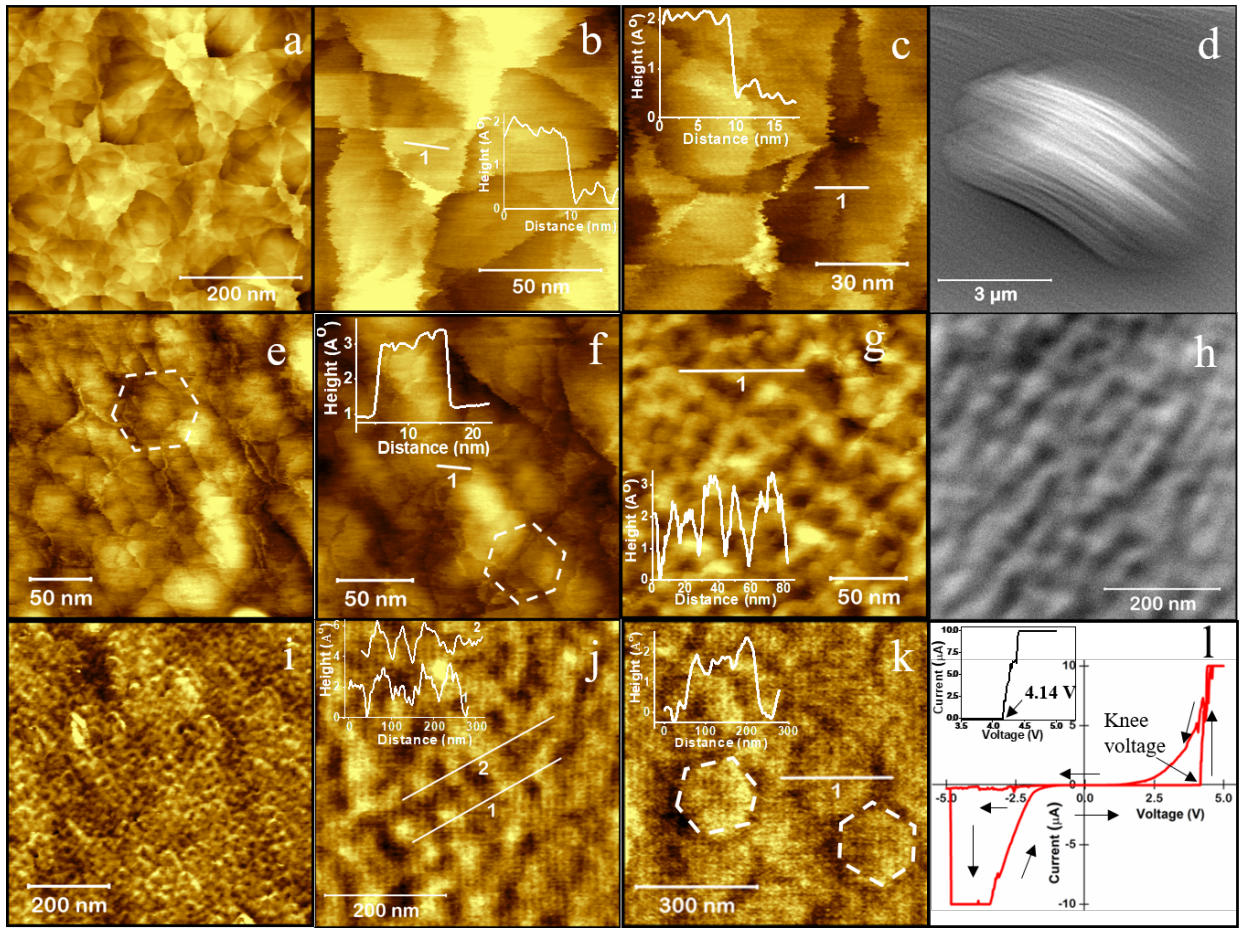

**Figure 2: Demonstration of one atom thin gold films on silicon and sapphire substrates:** In 2a we show an AFM image of gold films on sapphire after heat treatment at 475°C for 60 s. 2b and 2c are higher magnification images of regions in 2a. AFM height measurements along the white lines show 1.5 Å – 2.0 Å thinness confirm that these are one atom thin films. 2d is a lower magnification SEM image of a region showing a stack of thin gold films on silicon. In 2e we show an AFM image of a stack of thin gold films on silicon with clearly defined morphologies, after heat treatment at 475°C for 10 s. In Figure 2f we show a higher magnification AFM image of a region in 2e of thin films with well-defined morphologies. Height profile measurement along the white line 1 shows a film thinness of ~ 2 Å confirming that it is one atom thin. The white dotted lines in 2e and 2f highlight the hexagonal shape of these films. In 2g we show a region on the silicon substrate where we observed a highly ordered 2-dimensional array of gold structures. Height profile measurement (~ 2 Å) confirmed the presence of periodic one atom thin arrays. 2h is an SEM image of the same film confirming the 2-dimensional arrays observed by AFM in 2g. In 2i we show an AFM image of a region of the gold film on sapphire that was heat treated to 475°C for 60 seconds. The image shows a honeycomb type structure. A higher magnification image in 2j shows a one atom thin array as per the height profile measurement (~ 2 Å). In 2k we show an AFM image of an ordered array of gold film/discs that are hexagonal in shape. Height profile measurement across one such disc (~ 2 Å) confirmed that it is one atom thin. In 2l, we show a typical I/V spectrograph of the 2-dimensional gold films on silicon showing a very high *turn on* voltage of 4.14 V. The inset in the top left quadrant is a higher magnification plot of the I/V curve around the *turn on* voltage region.

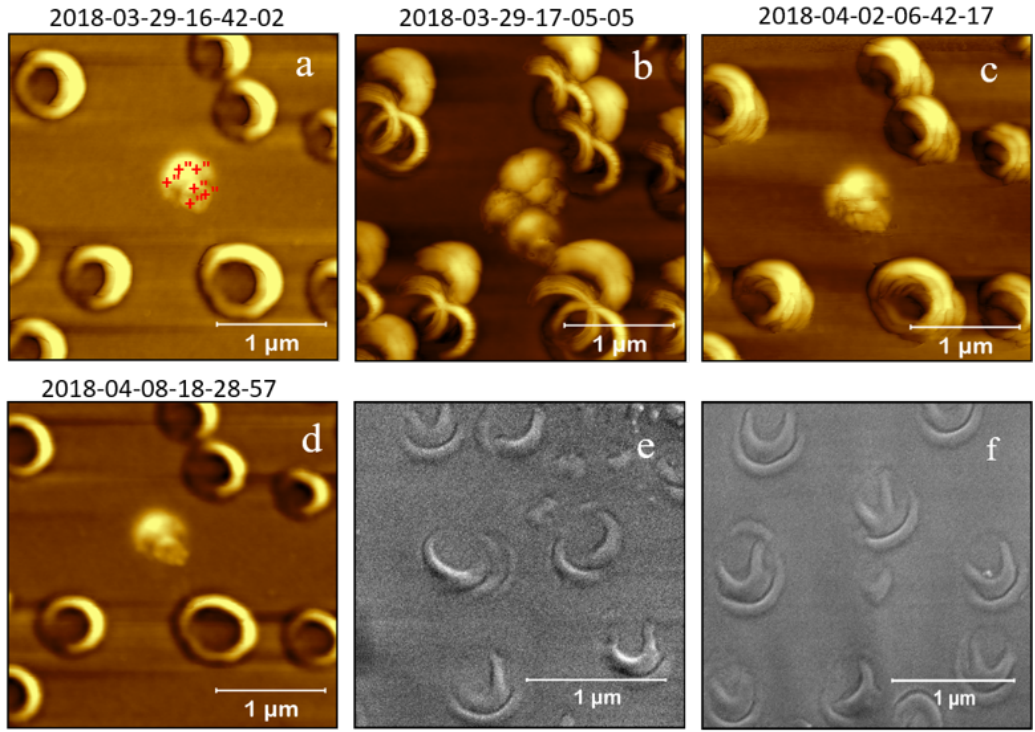

**Figure 3: Demonstration of unusual nanoring phenomena:** a-d are AFM images of the same sample field over a period of time. 3a shows an AFM topographic image of the gold nanoring structures, before being subjected to the applied force of 100 nN. The location of applied force is marked by the red ink **(+)** marks in 3a. The time and date of experiment is marked on the figure. In 3b, we show the same image field, immediately after the application of the force. It is clear that application of the force generated several *daughter-ring* replicas from each *mother-ring* structure. 3c shows the process of re-assembly in the intermediate stage. In 3d, all the *daughter-ring* replicas have completed their migration back to their *mother-ring* structure and re-assembled to create the original shape and orientation. The final image 3d is remarkably identical to that of the starting image 3a. In 3e and 3f we show the backscattered electron images of nanoring *mother-ring/daughter-ring* structures. The generation of several *daughter-ring* replicas from each *mother-ring* structure due to the application of force is observed.

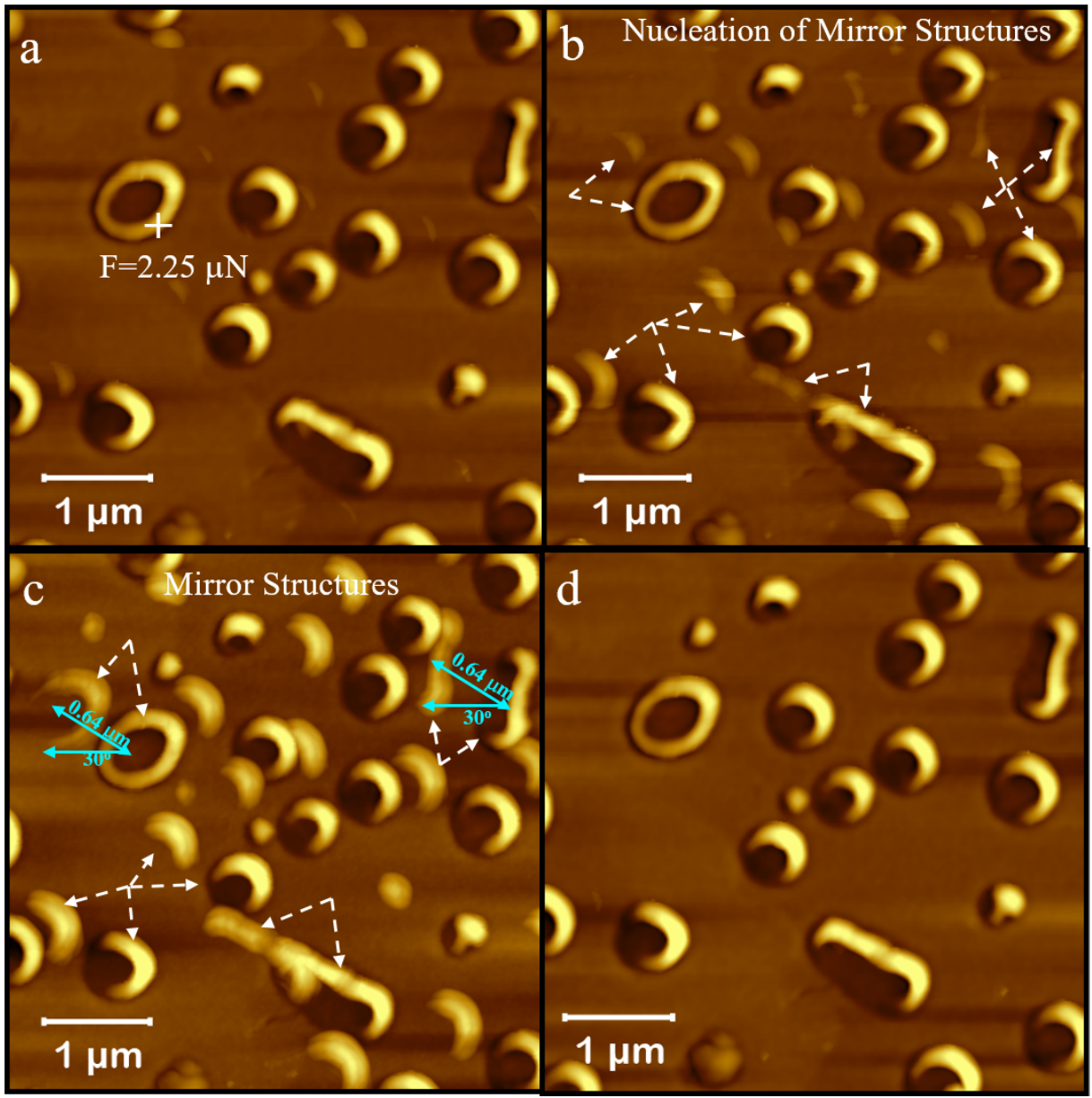

**Figure 4: Observation of the force response *mirroring* behavior in nanorings.** 4a is an AFM image of nanorings prior to the application of a point force of 2.25 µN for 120 s at the white (+) mark on a nanoring. 4b is an AFM image of the same field immediately after the application of the force. It shows the nucleation of the *mirror* structures, as indicated by the dotted white lines with arrows. 4c is an AFM image the same region after ~30 minutes. It shows well-formed, one *mirror* nanoring structure for every nanoring in the image field. All the *mirror* structures are aligned with each other at an angle of 30º to the horizontal plane. Each *mirror* structure is located at ~0.64 µm from its *mother* nanoring. 4d is an image of the same field after approximately an hour and no *mirror* rings were observed.

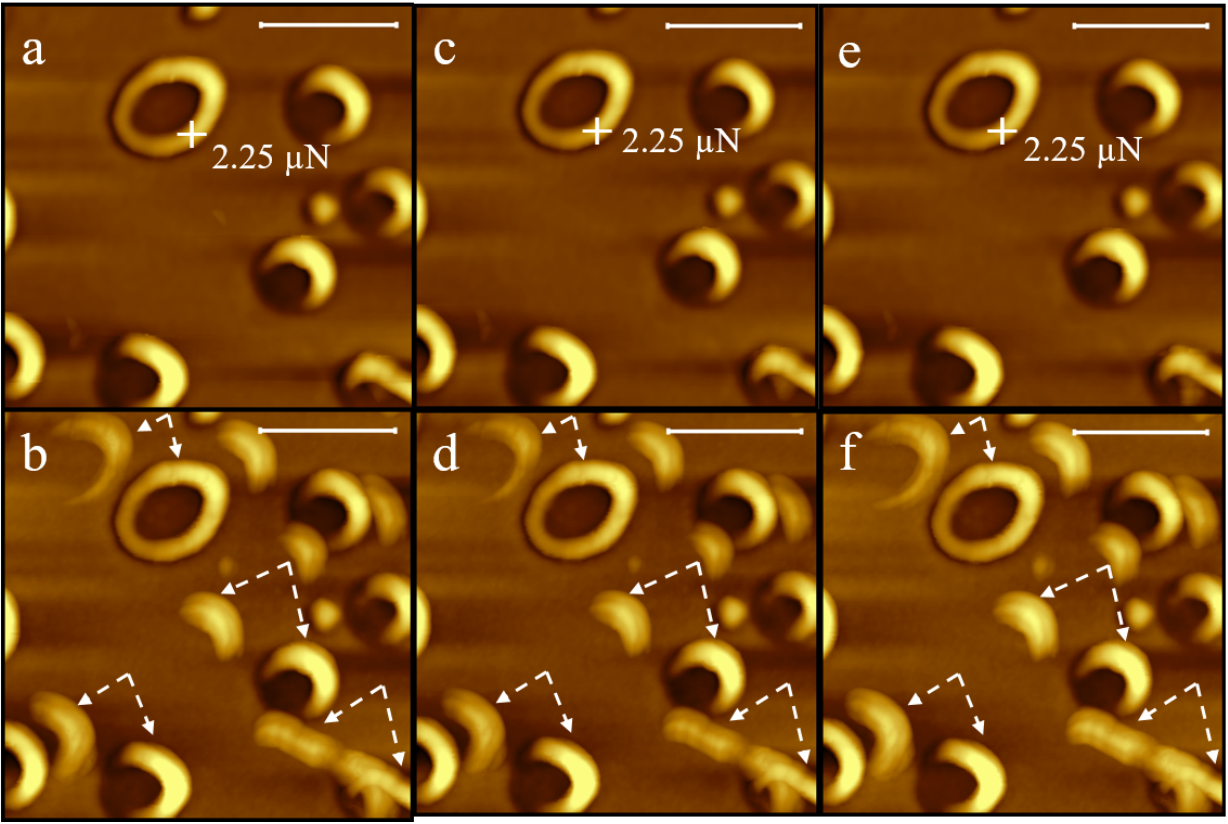

**Figure 5: Demonstration of the highly repeatable force response *mirroring* behavior of nanorings:** In 5a we show the AFM of a field of nanorings prior to the application of 2.25 μN for 120 s at the white (+) mark on a nanoring. In 5b we show the same image field after the application of the force in 5a and the well-formed *mirror* nanoring structures and their associated *mother* rings are indicated by the dotted white lines with arrows. In 5c we show the AFM image of the same field after approximately an hour and the *mirror* structures have disappeared. The same point force was applied at the same location indicated by the white (+) mark. In 5d we show the AFM of the same field after the application of the force and we can see the *mirroring* phenomenon seen in 5b reproduced. In 5e we show the AFM image of the same field after approximately an hour and the *mirror* structures have again disappeared. We applied the same force at the same location indicated by the white (+) mark. In 5f we show the AFM image of the same field after the application of the force. The *mirroring* phenomenon observed in 5b and 5d was again precisely reproduced. The scale marker is 1μm.

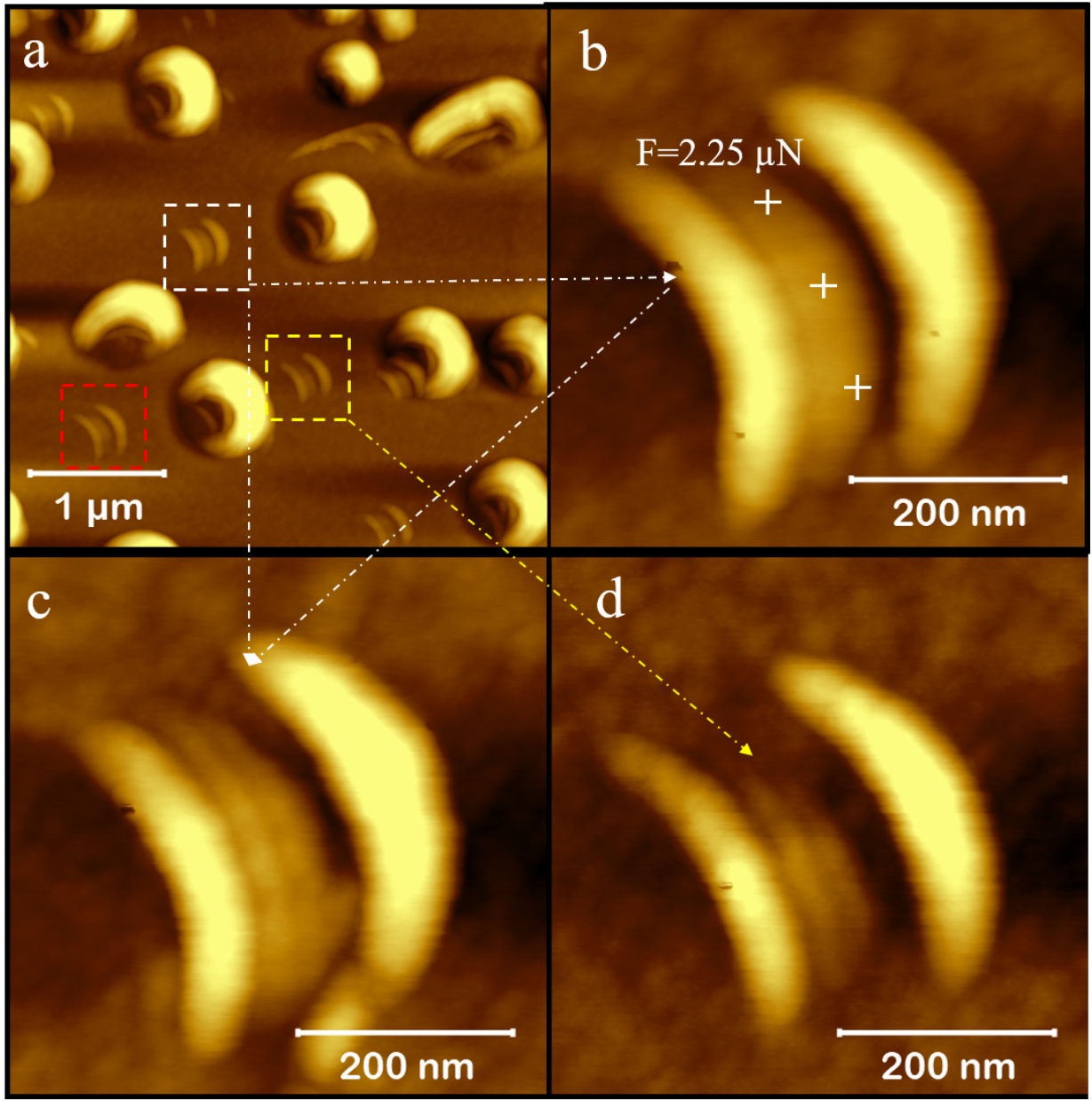

**Figure 6: Observation of the force response *mirroring* behavior in the *mirror* structures.** In 6a we show an AFM image of nanorings and their corresponding *mirror* structures. In 6b we show a higher magnification AFM image of the *mirror* rings within the white dotted square in 6a. We applied a force of 2.25 μN for 120 seconds to the middle *mirror* ring in three locations indicated

by the white (+) marks in 5b, in a sequential manner. In 6c we show the AFM image the *mirror* rings shown in 6b, immediately after the application of the force. We observed that the middle *mirror* ring has split into two identical rings. In 6d, we show the higher magnification image of the *mirror* rings within the yellow dotted square in 6a, after the application of the force shown in 6b. The middle *mirror* ring has split into two identical rings exhibiting a *mirroring* response to the force response shown by the *mirror* ring in 6c.

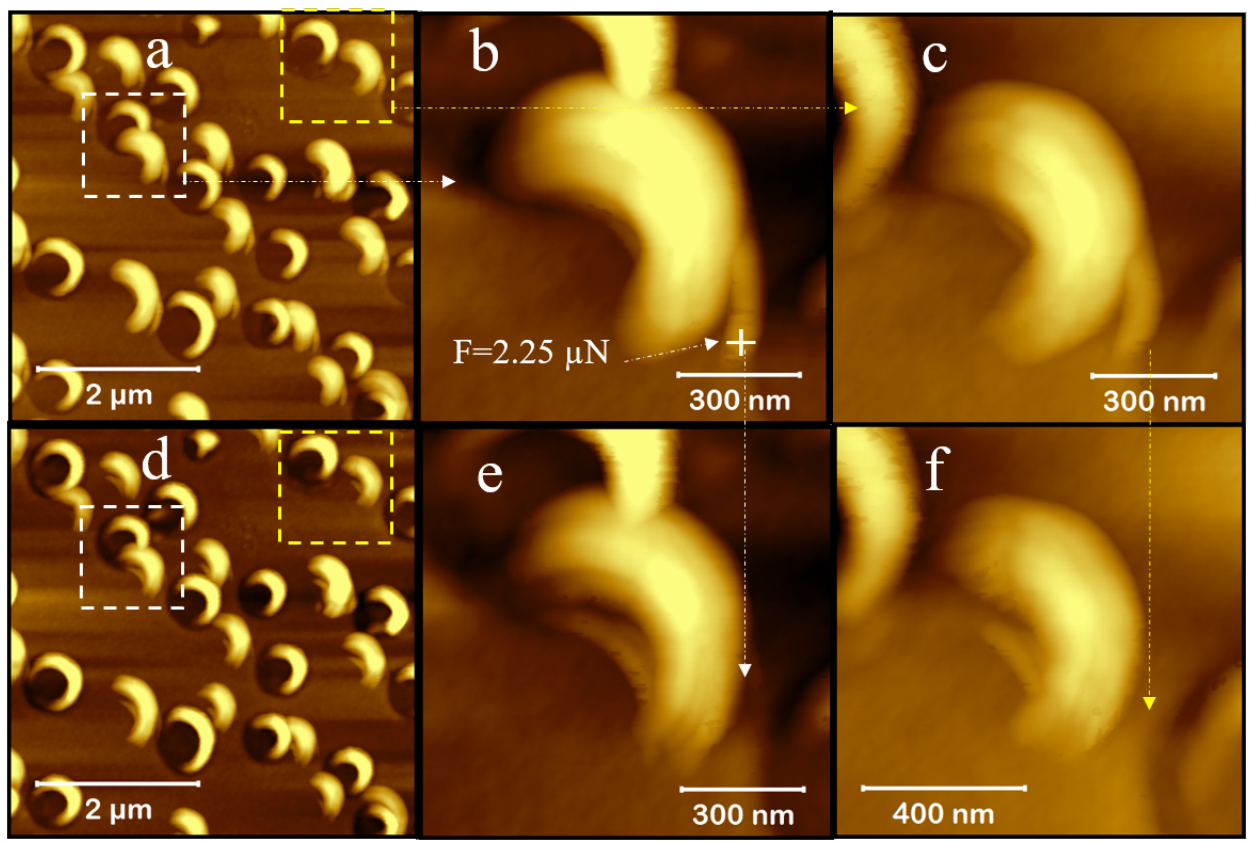

**Figure 7: Observation of the force response *mirroring* behavior in the *mirror* structures.** In 7a we show the AFM image of nanorings and their *mirror* ring structures. In 7b we show a higher magnification image of the *mirror* ring within the white dotted square in 7a. In 7c we show the higher magnification image of the *mirror* ring within the dotted yellow square in 7a We applied a point force of 2.25 µN for 120 seconds to the *tail* in the *mirror* ring at the location indicated by the white (+) mark in 7b. In 7d, 7e and 7f, we show the images of the fields shown in 7a, 7b and 7c, after the application of the force. We observed that application of the force to the *tail* in the *mirror* ring in 7b resulted in its disappearance of *tail* as shown in7e. This behavior (i.e. disappearance of the *tail*) was simultaneously mirrored by all the *mirror* rings in the image field as observed in 7d and 7f.

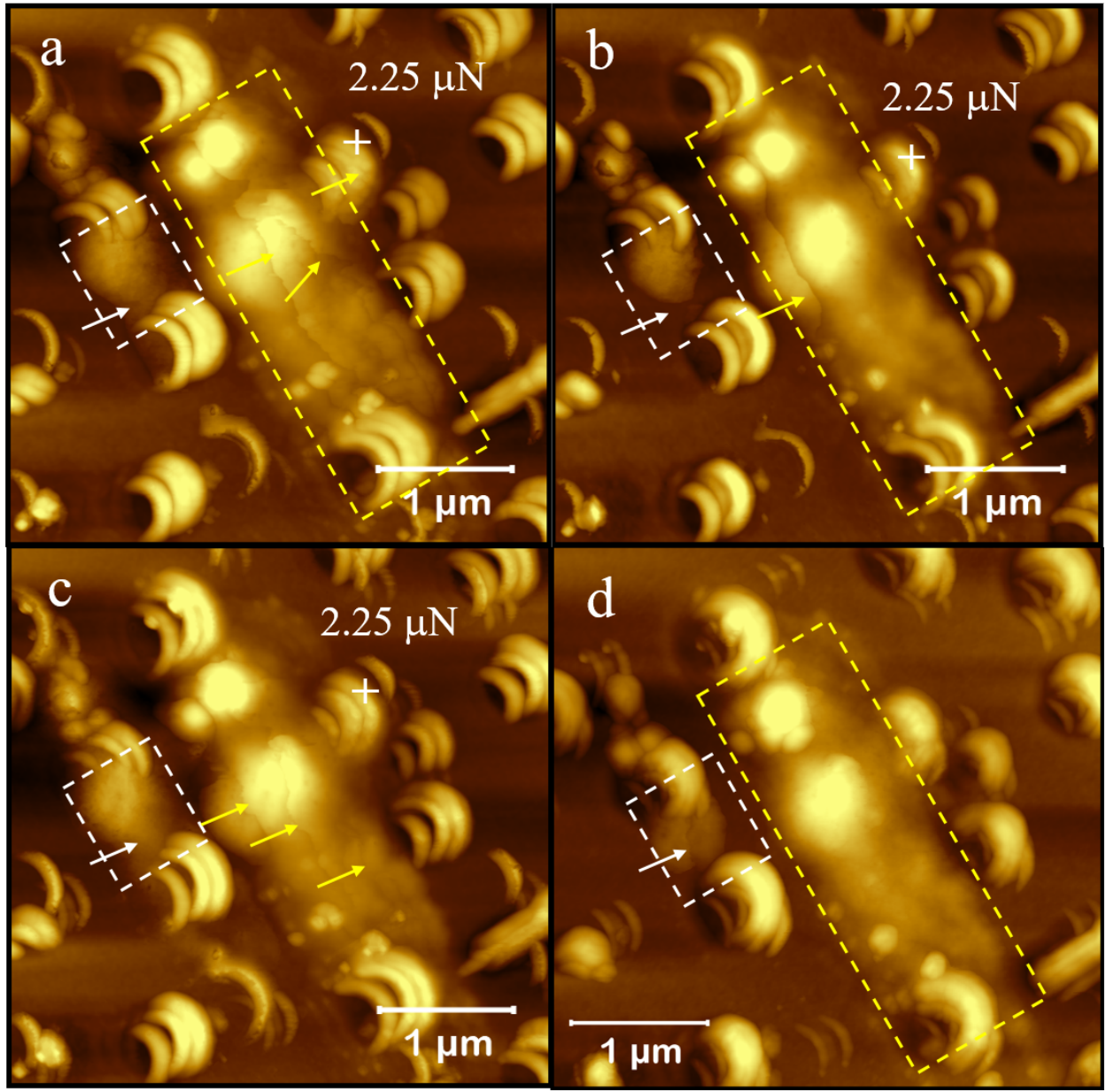

**Figure 8: Room temperature *liquid like* gold.** In 8a we show the AFM image of several gold nanorings, a large area of gold film in the middle within the yellow dotted rectangle and a smaller film to its left within the white dotted rectangle. The yellow arrow lines show the liquid wave fronts on the large film and on the nanoring with the white (+) mark. The white arrow line shows the liquid wave front on the smaller film. We applied a point force of 2.25 µN for 120 seconds to the nanoring at the location indicated by the white (+) mark. In 8b we show the AFM image of the same field and we observed that the liquid wave fronts on the large film in the middle have coalesced together into one wave front and moved forward as indicated by the yellow arrow. The smaller liquid film within the white dotted rectangle has de-wetted and receded as indicated by the white arrow. We again applied the same force at the same location. In 8c we show the AFM image of the same field after the force was applied. We observed that more liquid wave fronts have been formed on the large film in the middle as indicated by the yellow arrows. The smaller film within the dotted white rectangle has spread and the wave front has moved forward. We again applied the same force at the same location. In 8d we show the same image field after the application of the force. We observed that the liquid wave fronts on the large film in the middle have coalesced and now cover the entire film. The smaller liquid film within the dotted white rectangle has de-wetted again and receded as indicated by the white arrow.

# Supplementary Materials:

**Synthesis of One Atom Thin, Two-Dimensional Gold Films and Their Novel Properties**

Sudhir Kumar Sharma,[1] Renu Pasricha,[2] James Weston,[2] Florian Stumpf [3], Thomas Blanton[4] and Ramesh Jagannathan[1*]

Correspondence to: rj31@nyu.edu

**Figure S1.**

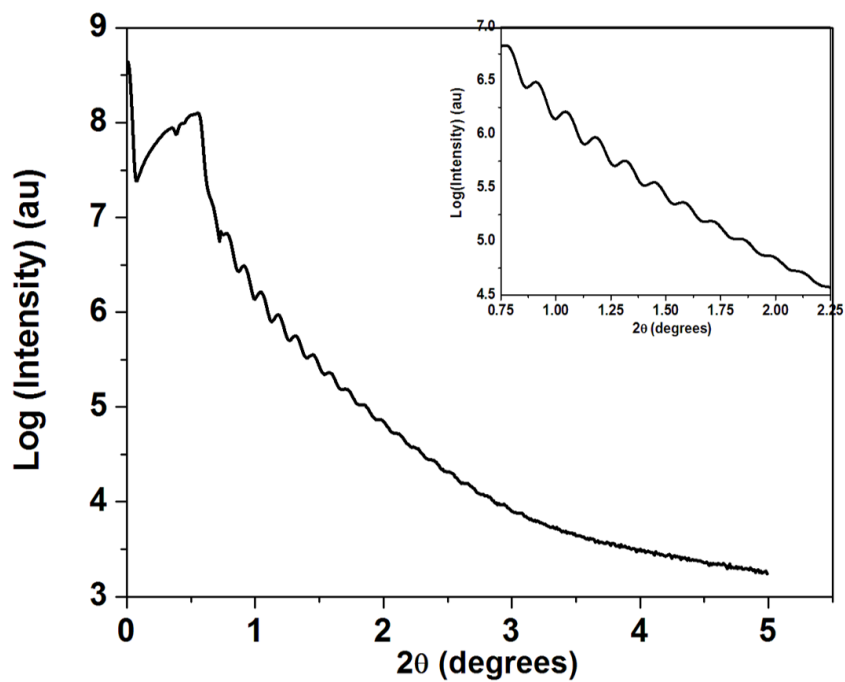

**Figure S1: X-ray reflectivity (XRR) pattern for as-deposited gold film on (0001) sapphire substrate:** Measured film thickness, 66 nm. Inset: selected range of XRR pattern. [Data collected at room temperature using a Bragg-Brentano diffractometer, Cu X-ray tube, reflection mode geometry]

Figure S2.

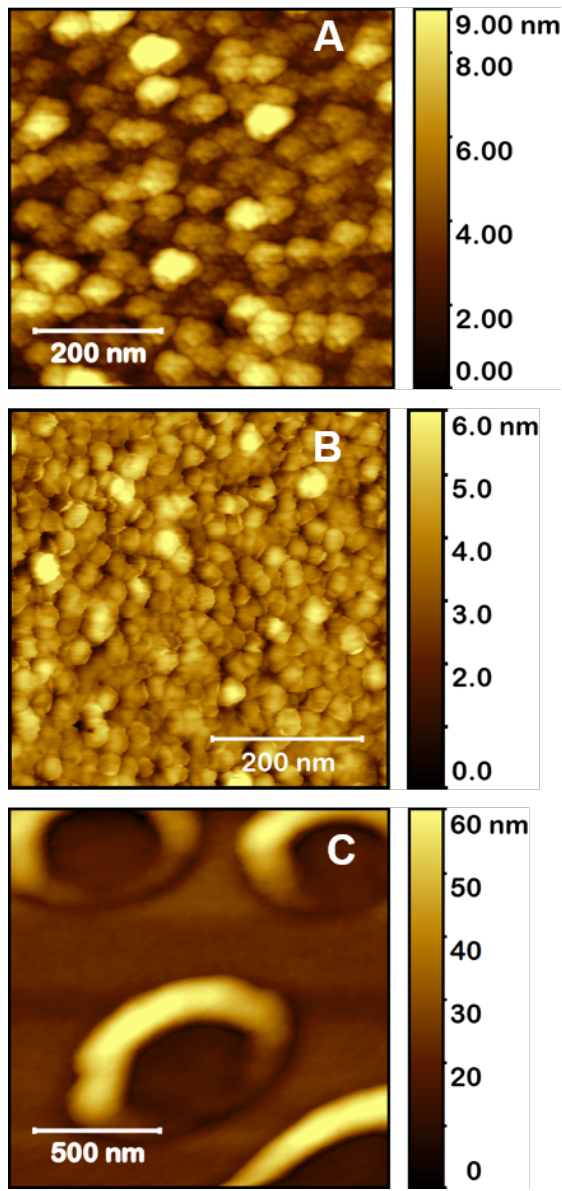

**Figure S2: Effect of heat treatment temperature on gold thin film on sapphire**: In A we show that subjecting the film to 350°C for 5 seconds results in solid-state de-wetting of the film and formation of crystallites. In B we show a similar effect at 425°C for 5 seconds. In C we show that subjecting the gold film to 475°C for 5 seconds results in the formation of nanorings.

**Figure S3.**

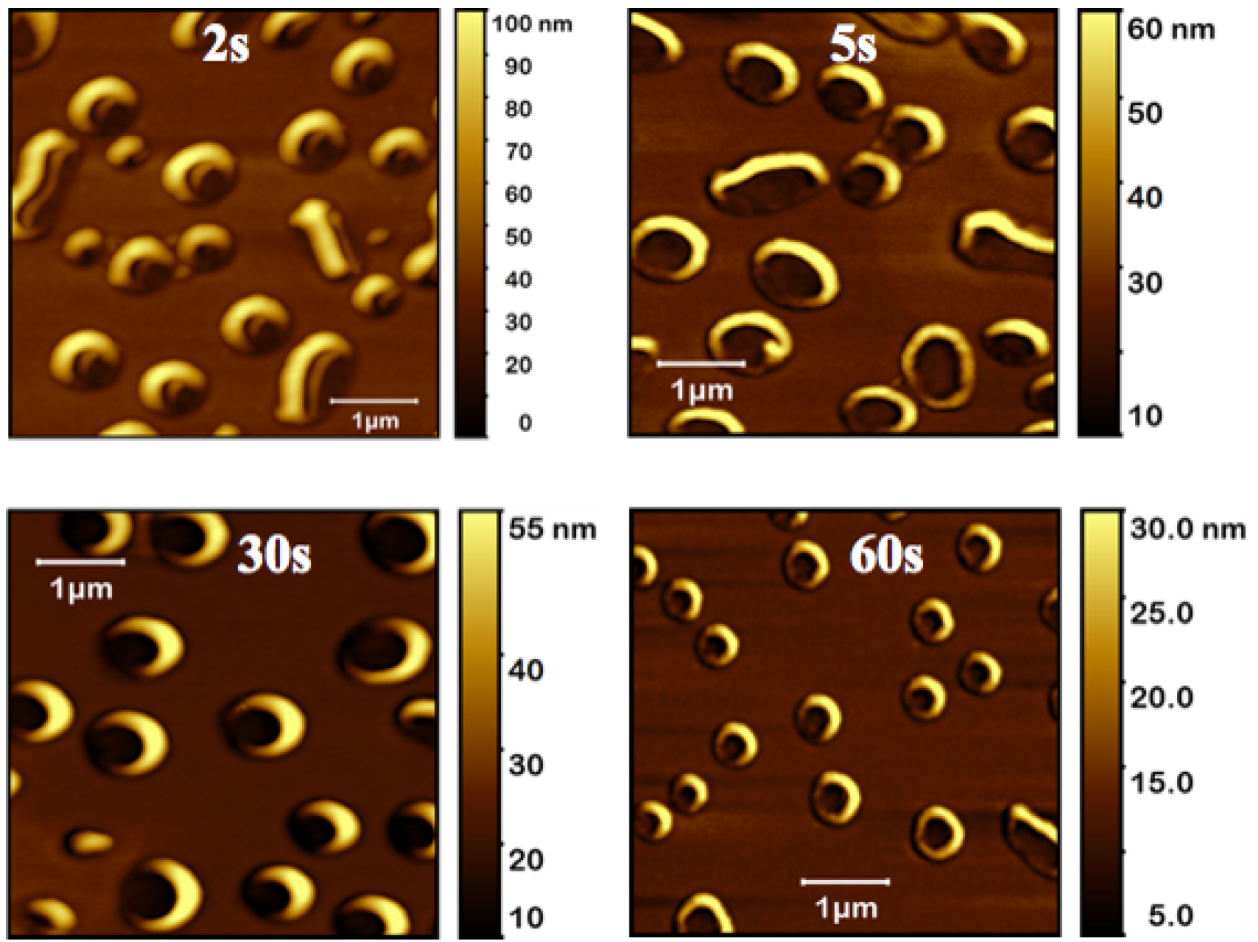

**Figure S3: Effect of heat treatment time on sapphire:** Gold nano-ring structures were observed when the 66 nm gold film on sapphire was treated to 475°C for 2s, 5s, 30s, and 60s, respectively.

**Figure S4.**

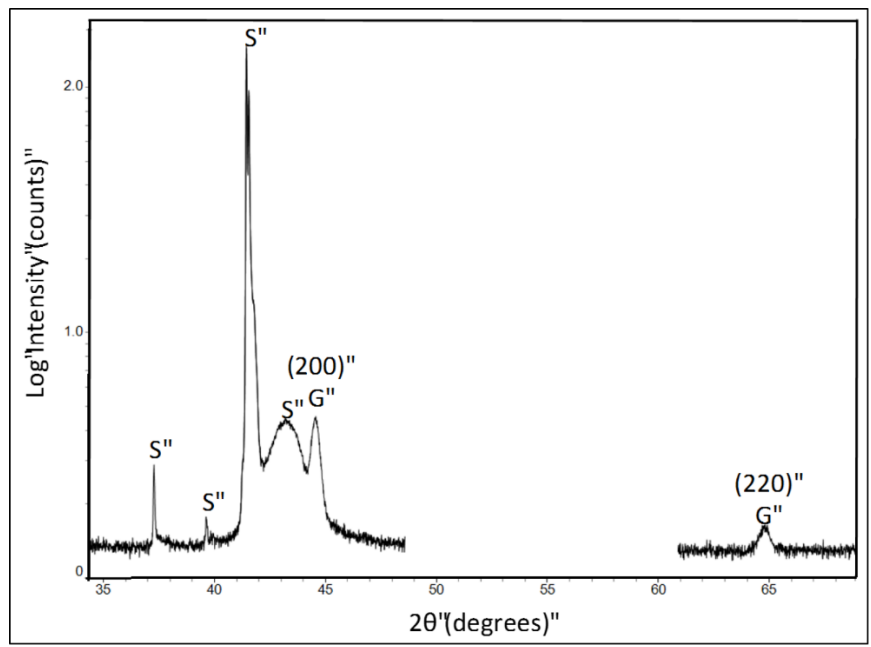

**Figure S4: Identification of the gold nanorings by X-ray diffraction on sapphire:** Selected range X-ray diffraction (XRD) patterns for gold on (0001) sapphire substrate after exposing sample to 475 °C for 60s. Diffraction peaks marked S are due to the sapphire substrate. The peaks marked G are due to gold corresponding to the (200) and (220) diffraction peaks. [Data collected at room temperature using a Bragg-Brentano diffractometer, Cu X-ray tube, reflection mode geometry]

**Figure S5.**

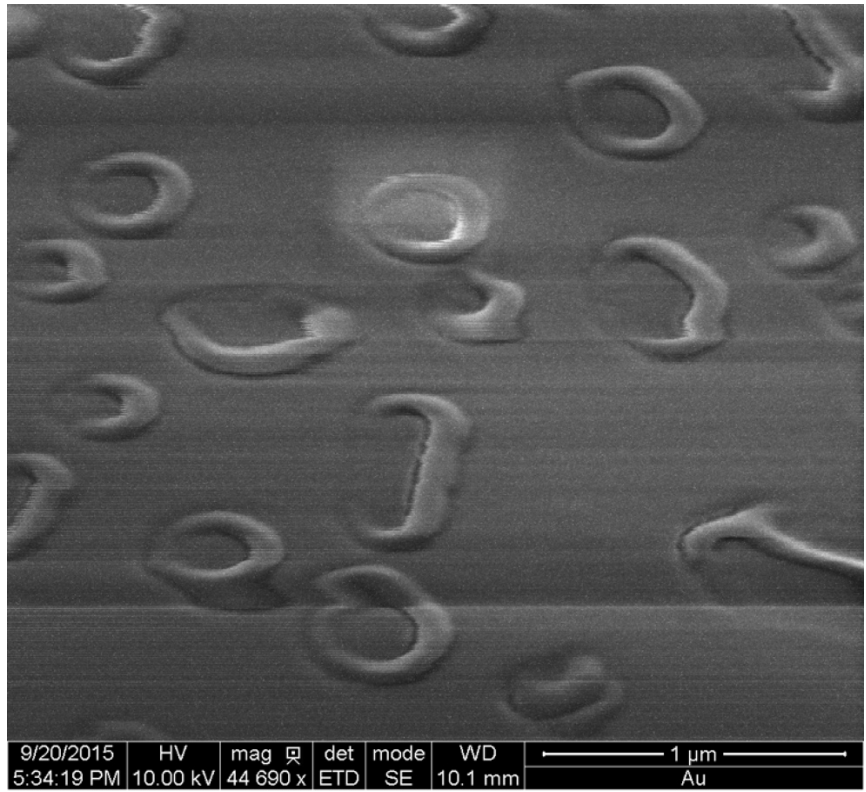

**Figure S5: SEM imaging of gold nanoring sapphire:** The susceptibility of the gold nano-rings to e-beam exposure during SEM imaging.

**Figure S6.**

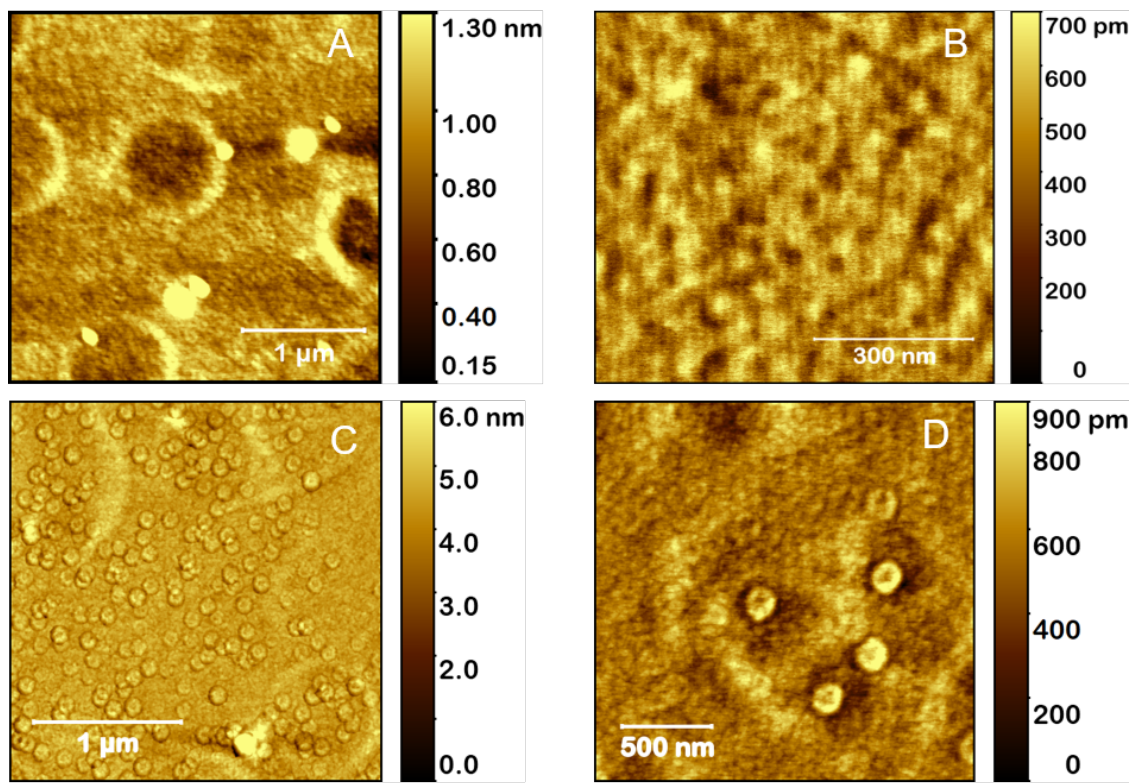

**Figure S6: Membrane-like porous gold films sapphire:** In A we show a (3μm x 3μm) region of the sapphire substrate covered with membrane-like gold mesh/film. B is a higher magnification image of a region shown in A. In C and D we show two regions of the sapphire substrate covered with hexagonal shaped one atom thin gold structures.

Figure S7.

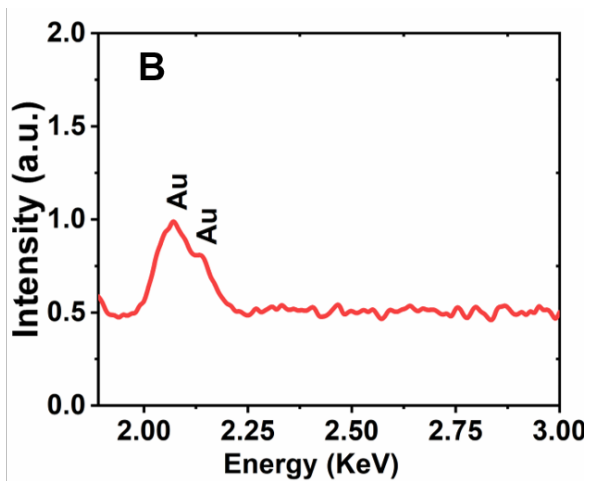

**Figure S7: EDS analysis of gold film on silicon:** We show the EDS spectrum obtained from the film on silicon confirming that these are gold nanostructures.

Figure S8.

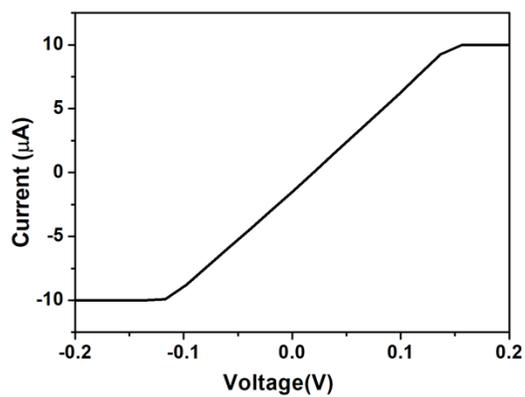

**Figure S8: IV Spectra for non-heat-treated gold films on silicon:** IV ohmic response for the non-heat-treated gold film.

**Figure S9.**

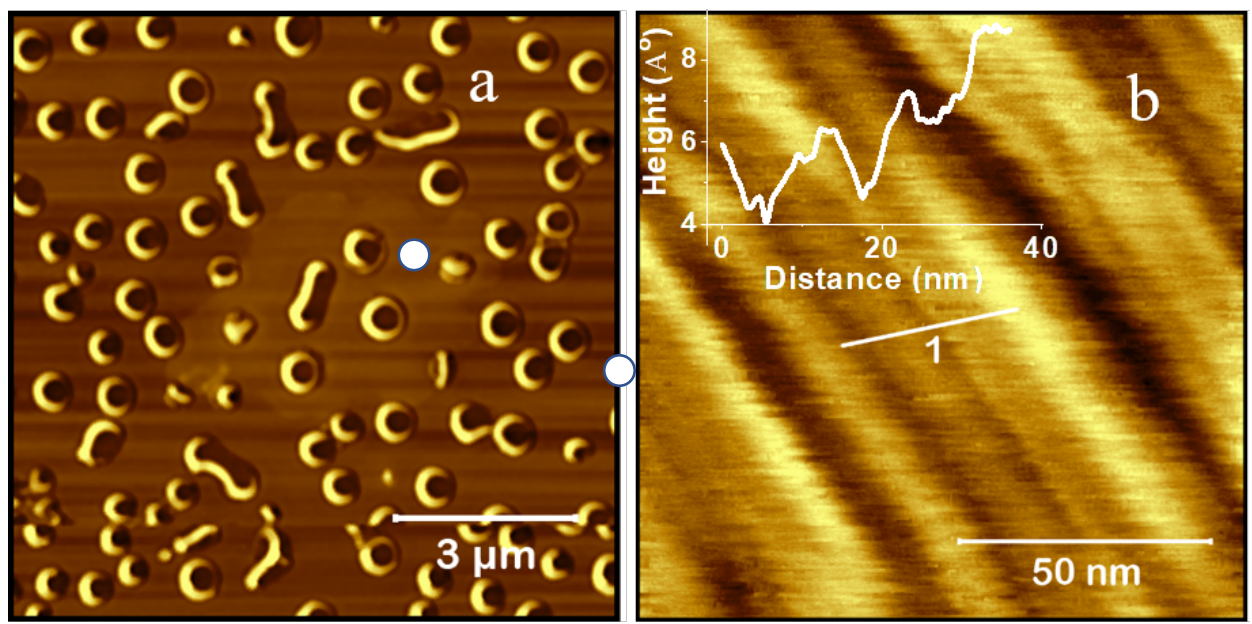

**Figure S9: One atom thin gold films on sapphire:** In "a" we show the AFM image of the nanoring structures. "b" is the high-resolution image of the area marked by the white circle in "a". It shows a stack of one atom thin gold films on the sapphire substrate adjacent to the nanoring structures.

**Figure S10.**

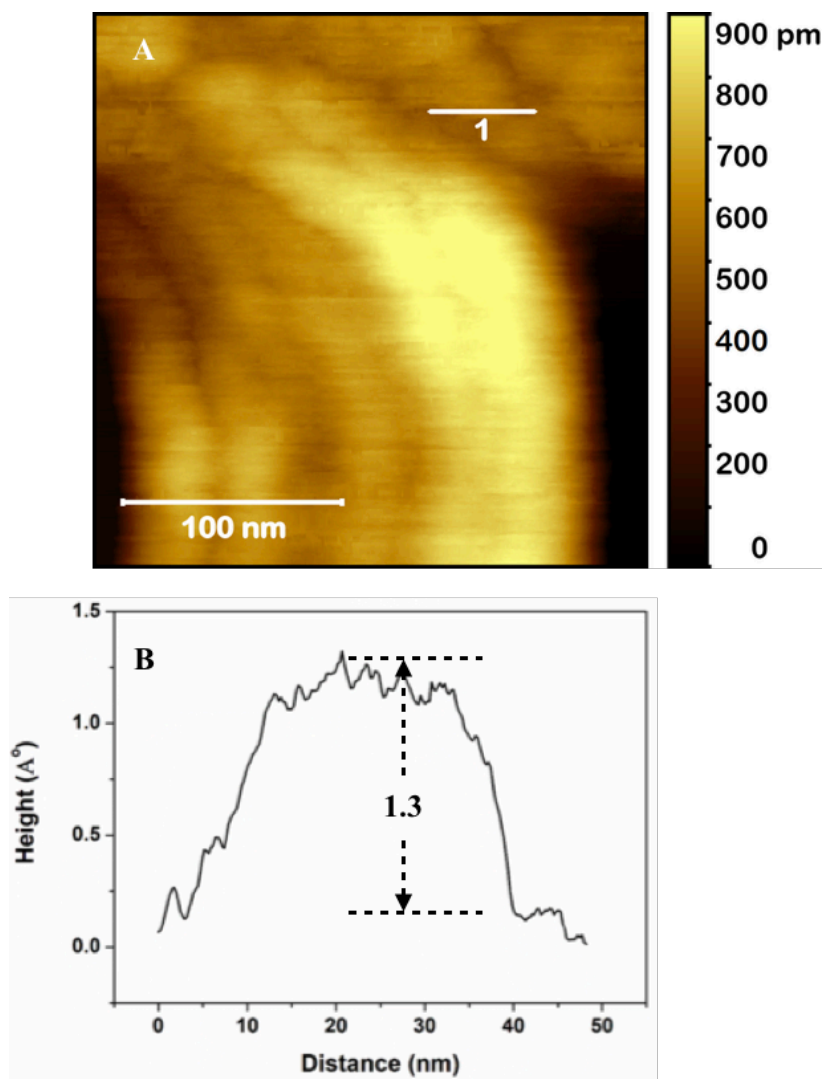

**Figure S10: One atom thin gold film on sapphire:** AFM image of the nanoring and the region adjacent to it showing the presence of thin gold films. Height measurement show the film to be 1.3 Å, indicating a one atom thin film.

**Figure S11.**

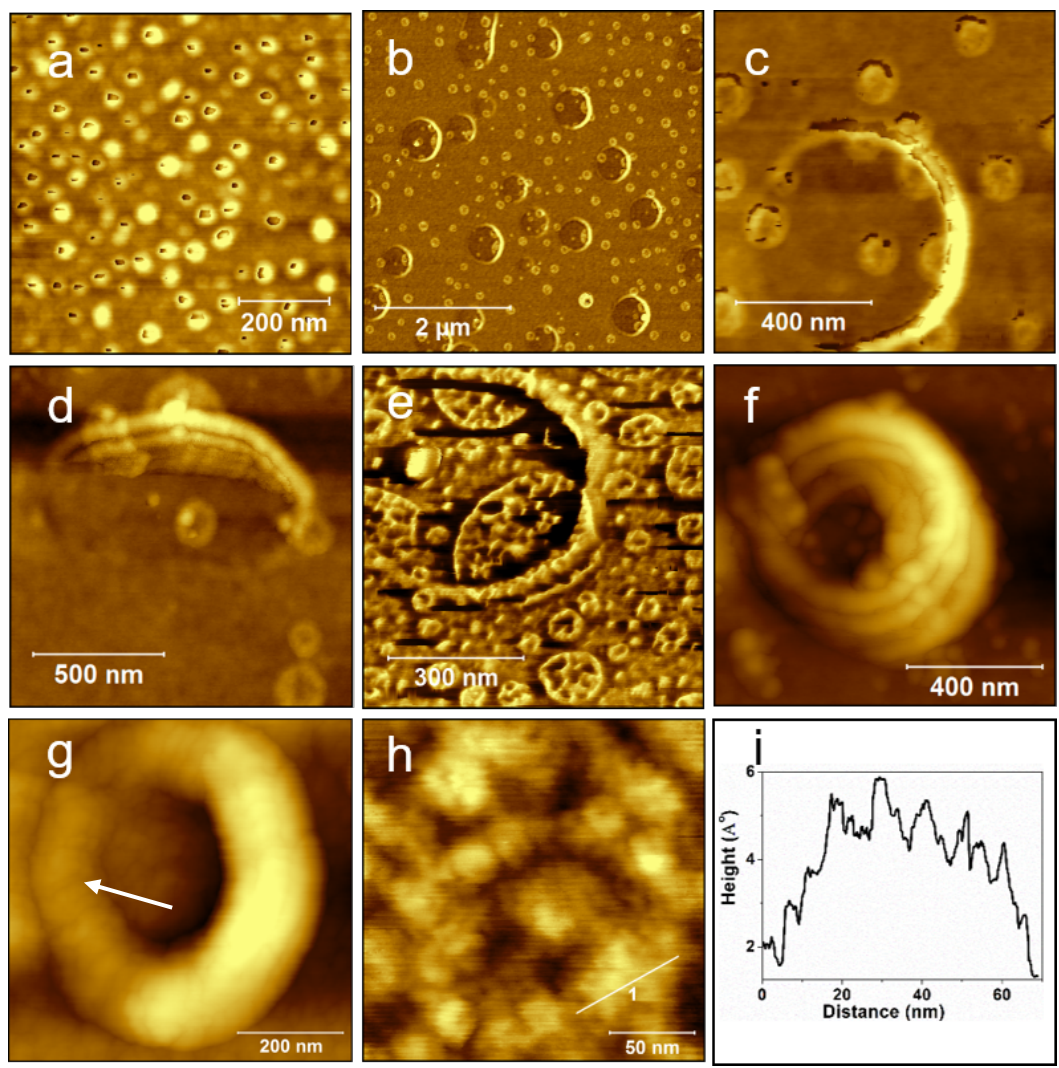

**Figure 11: Mechanism of Nanorings -** In this Figure, **a** shows several open dome-like structures, with their edges pinned to the sapphire substrate. The open-dome like structures coalesce first to form larger, arc-like structures (Figure S11b and S11c). In Figure S11d, we show the evolution of the vertical growth stacks of these arc-like structures, more clearly seen in the AFM phase image Figure S11e. In Figure S11f, we show the topographic image of an almost completed nanoring structure, made up of stacks of several thinner rings. Figure S11g is the image of a completed nanoring. High-resolution topographic AFM image of a selected area of the nanoring in S11g

(white arrow) is shown in Figure S11h. In Figure S11i, we measured the height profile across the smaller discs (building blocks) to be approximately 1.7 Å, implying that these discs are made up of one atomic layer of gold.

**Figure S12.**

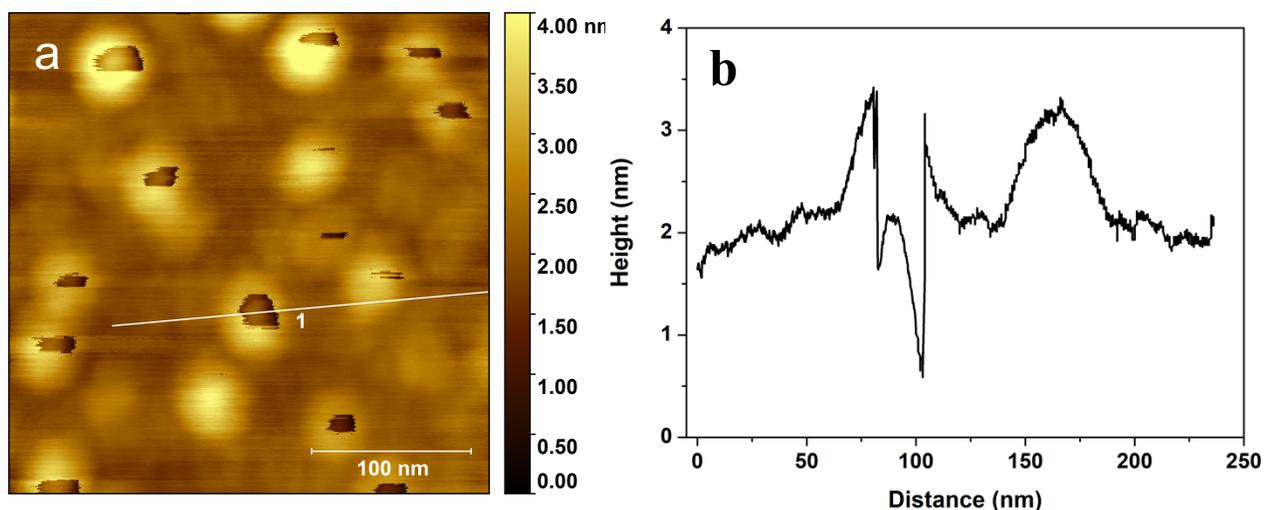

**Figure S12:**

**Demonstration of the "open dome" like structures:** S12a, is a higher magnification topographic image of the structures shown in Figure S11a. We measured the quantitative height profile across one of the open-dome like structures marked as 1 (Figure S12a). In Figure S12b, we show the height profile corresponding to the marked line 1 extracted from the Figure a. The open-dome like structures rise from the pinned edges to a height of about 2 nm over an axial distance of 50 nm on either side of the open top.

**Figure S13.**

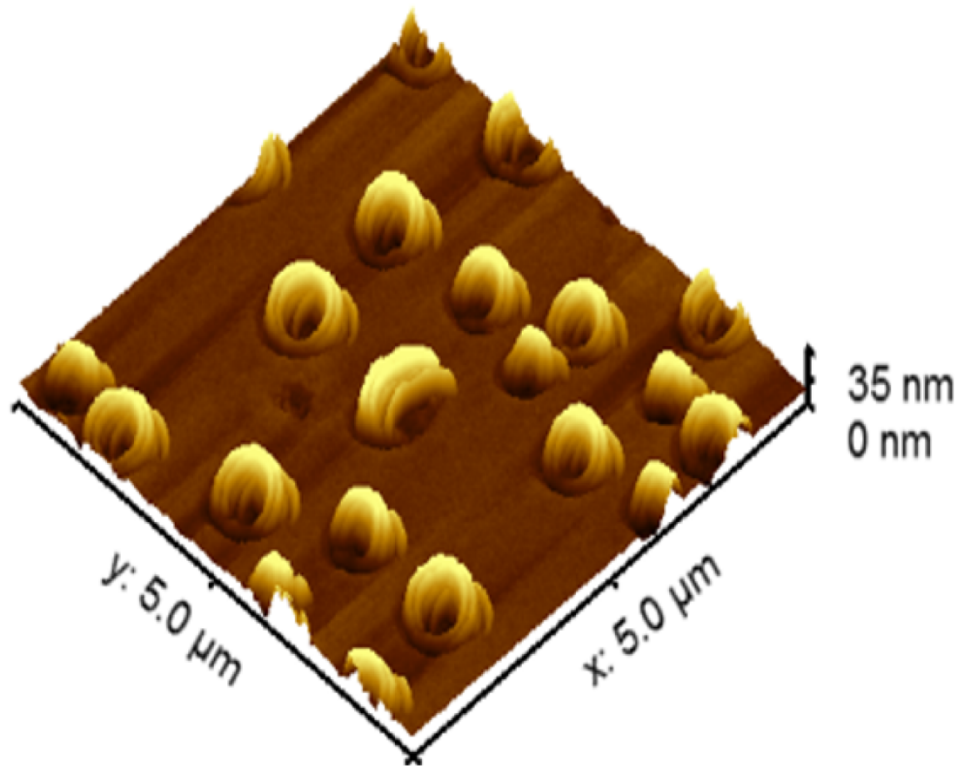

**Figure S13: 3-D representation of gold nanorings:** 3-D view of AFM data collected for the gold nano-ring structures demonstrate their asymmetric growth morphology. The nano-ring structure in the center of the image frame shows fine features of rapid growth, perpendicular to the substrate surface.

**Figure S14**.

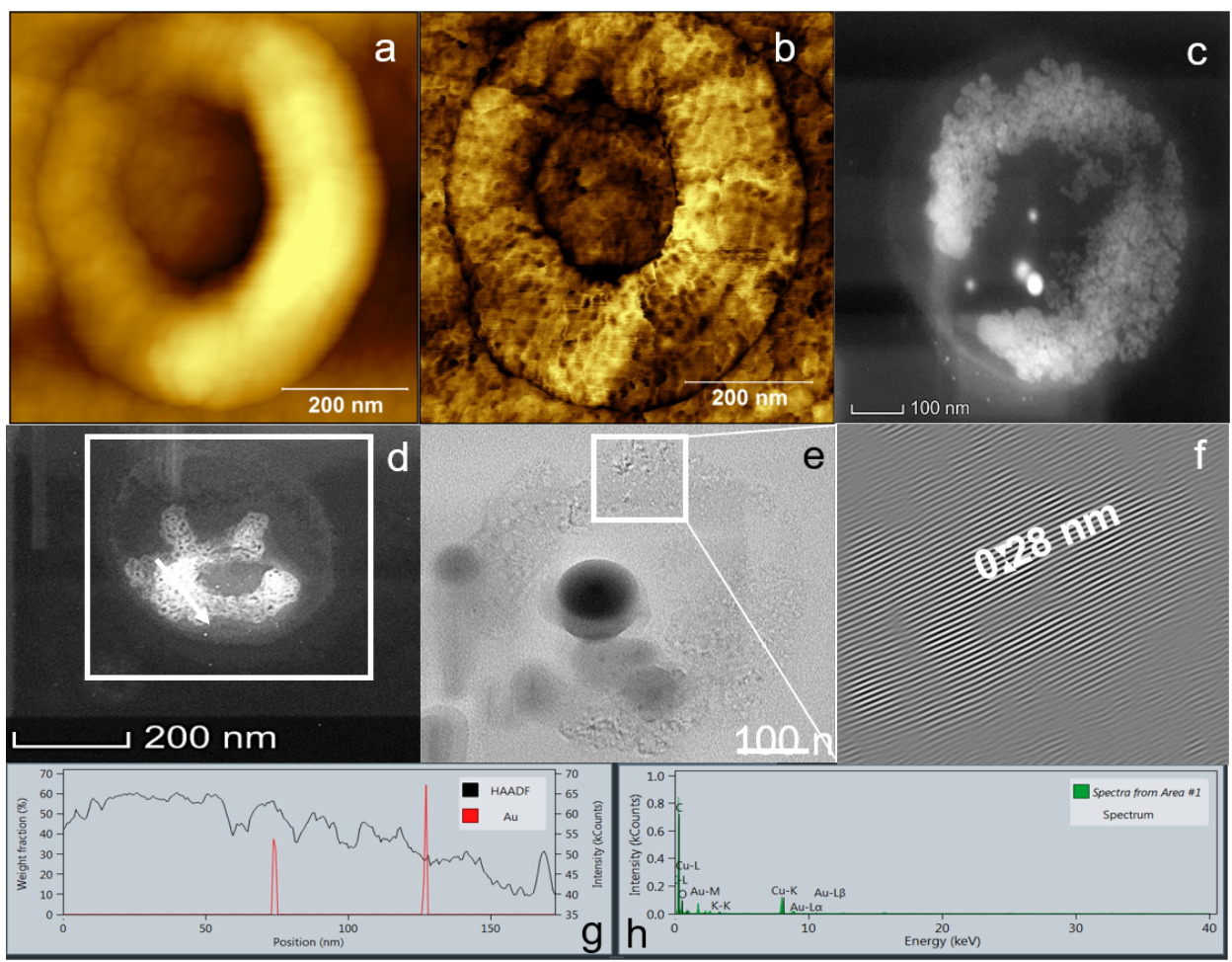

**Figure S14: Confirmation of gold nanorings**. a and b are AFM images of a typical nanoring structure. c and d show the STEM images of two different nanorings. e shows the high-resolution nanoring surface imaging in bright field mode. The direction of incident beam was normal to the specimen surface. f shows the lattice parameters of the HRTEM image, reconstructed using IFFT. Elemental X-ray spectral mapping (EDS) of Au shown in g and h, was used to confirm the location of Au in the nanoring shown in d (white line) confirming the presence of Au across the ring. h shows the integration of characteristic peaks of Au centered at 2.1 and 9.7 keV in EDS spectra.

**Figure S15**.

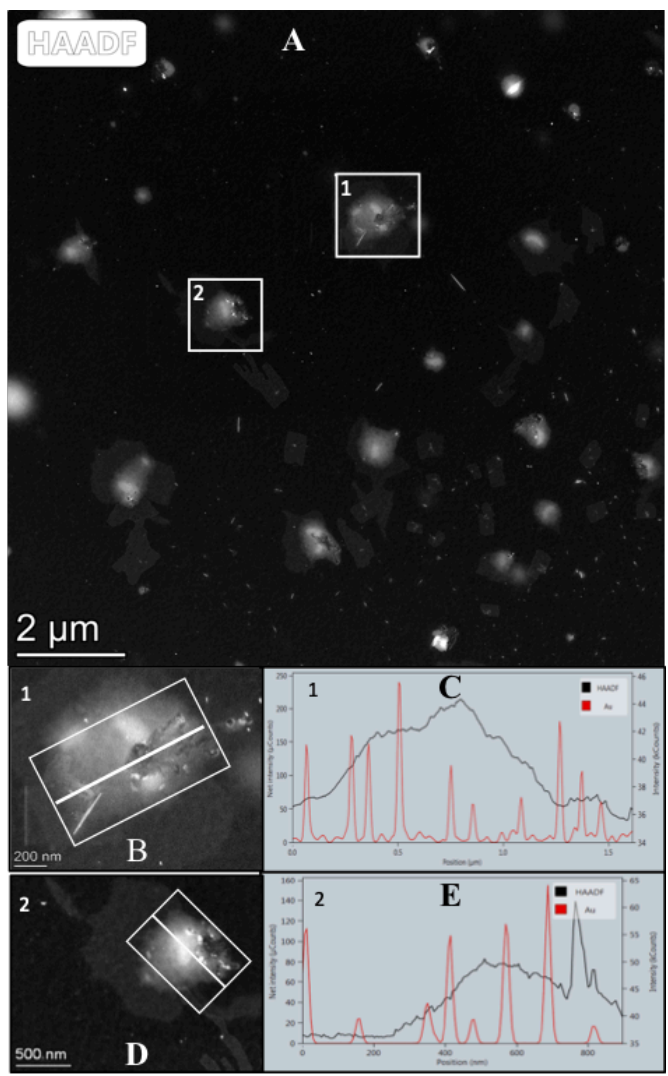

**Figure S15: Demonstration of the presence of gold across the entire area of the nanorings:** In A we show TEM images of several nanorings. The areas marked 1 and 2 of two nanorings were imaged at a higher magnification and shown in B and D. In C and E, we show elemental X-ray spectral mapping (EDAX) of Au across individual nanoring structures along the white line, in the boxes 1 and 2 respectively.

S14

**Figure S16.**

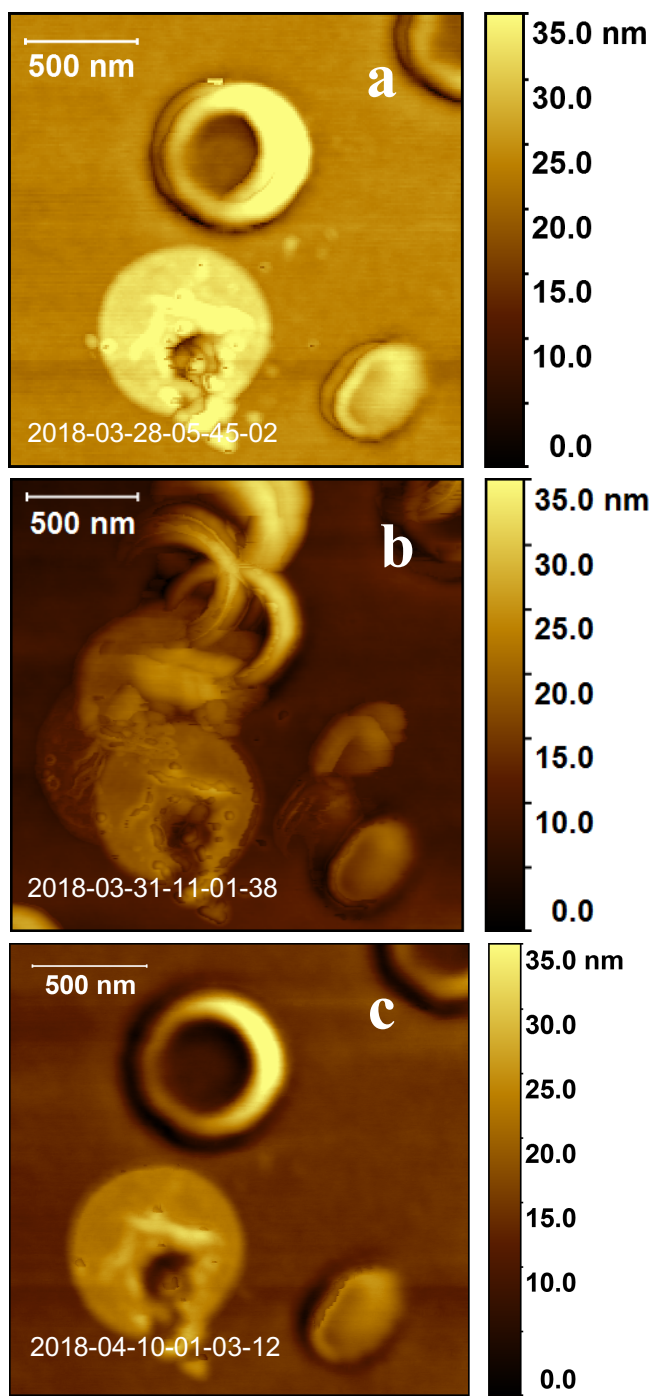

**Figure S16: Unique spatial memory register of *daughter* structures:** S16a shows gold structures before the application of the weak force. S16b is the image of the gold structures after the application of the force. It is clear that the fragment replicas from two of the "mother" structures significantly overlap with each other. S16C demonstrates the property of spatial memory and migration to their respective "mother" structures of the fragment replicas. The fact that they are in significant physical contact with each other did not prevent them from migrating back to their original "mother" structure.

**Figure S17.**

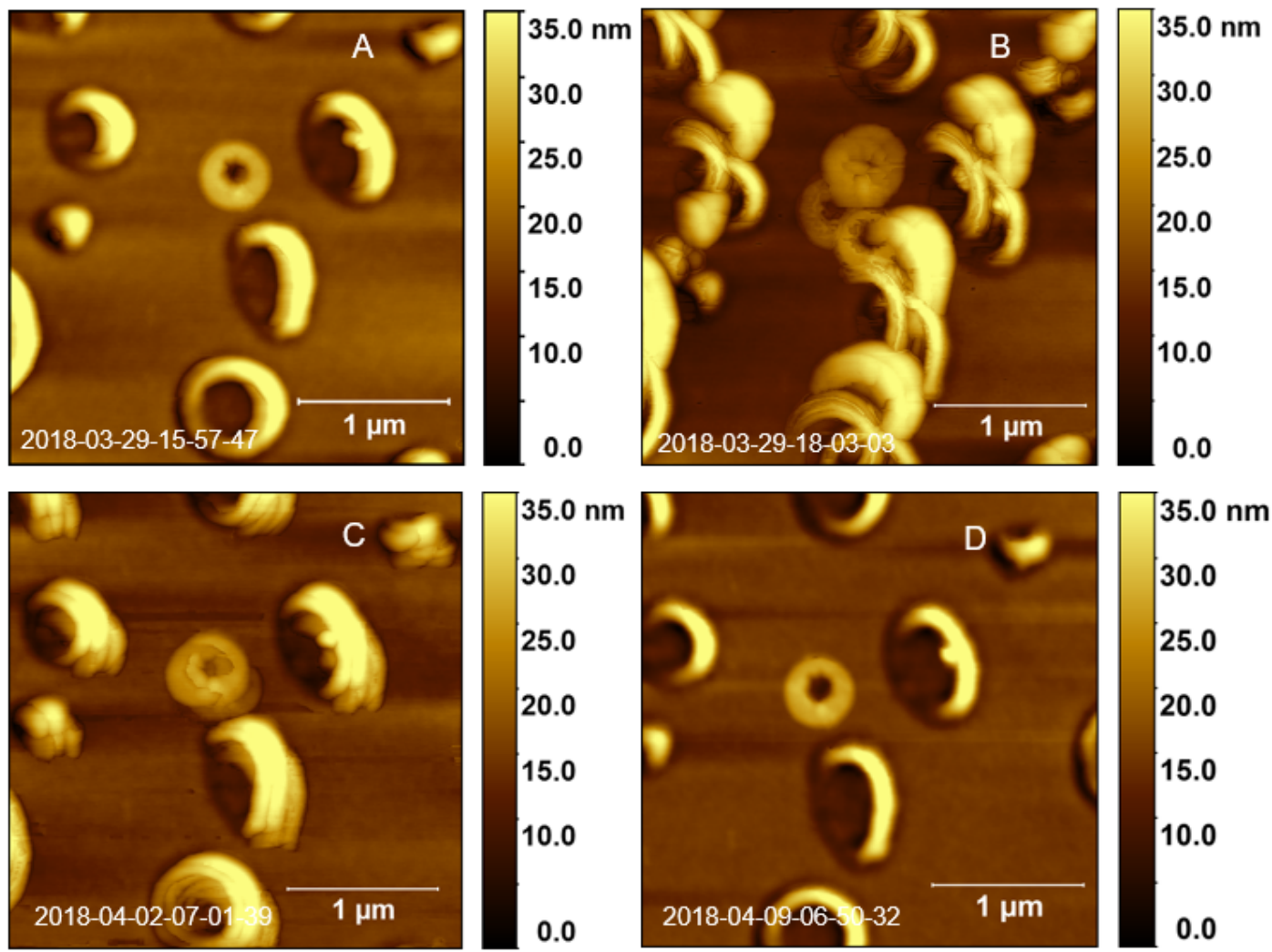

**Figure S17: Simultaneous observation of the "*memory register*" at a location 45 μm away from the point of application of the force, as marked in Figure 3**: In A we show an AFM topographic image of the gold nano-ring structures, before the application of the force. In B, we

S17

show the same image field, immediately after the application of the force. It is clear that application of the force generated several 'daughter' replicas from each 'mother' structure, similar to the observation made from Figure 3b. Over a period of time, these 'daughter' replicas migrated back to the original 'mother' structure location and assembled to re-create the original shape, as was also observed in Figure 3c and 3d. C shows the process of re-assembly in the intermediate stage. In D, all the 'daughter' replicas have completed their re-assembly process to create the original shape. The final image D is remarkably identical to that of the starting image A.

**Figure S18.**

**Experimental protocol to eliminate "tip/imaging artifact" in AFM measurements:**

Due to the presence of unique and unusual features observed in our experiments, we took extreme precautions to eliminate any "tip/imaging artifact" in the reported data. In general, artifacts are observed due to tip contamination with a foreign substance or a damaged tip. In either case, the effect would persist until the foreign substance is removed or replaced with the new tip. Also, an affected tip would generally result in a specific characteristic image. It is highly improbable that the same type of "tip artifact" would be observed when different tip/cantilevers are used to image different samples at different times.

In general, we followed the following protocols as our standard practices developed by the AFM imaging community.

1. Imaging the same region more than once, at several magnifications, scan speeds, scan angles (like 0º and 90º as shown below) and observe any inconsistencies between the images.
2. Examining for any potential inconsistencies and differences between trace and re-trace images.
3. Obtaining image data from different scan directions / scan angles of the same sample region
4. Repeating imaging with different samples.
5. Repeating imaging with different tip/cantilevers.
6. Using a "true" non-contact mode (NCM) imaging.

Moreover, at a logical level, the observed time dependent behavior of our experimental results (i.e. images) ruled out the presence of any persistent "tip artifacts" in the observed data. For example, in our imaging experiments with nanorings, using the same tip/cantilever system, we observed the appearance and disappearance of the so-called *daughter* structures over a finite time period, in a repeatable fashion. The same phenomenon was observed with different nanoring samples using different tips/cantilevers over a two-year period. The backscattered electron imaging of the *daughter* fragments further corroborated the AFM observations.

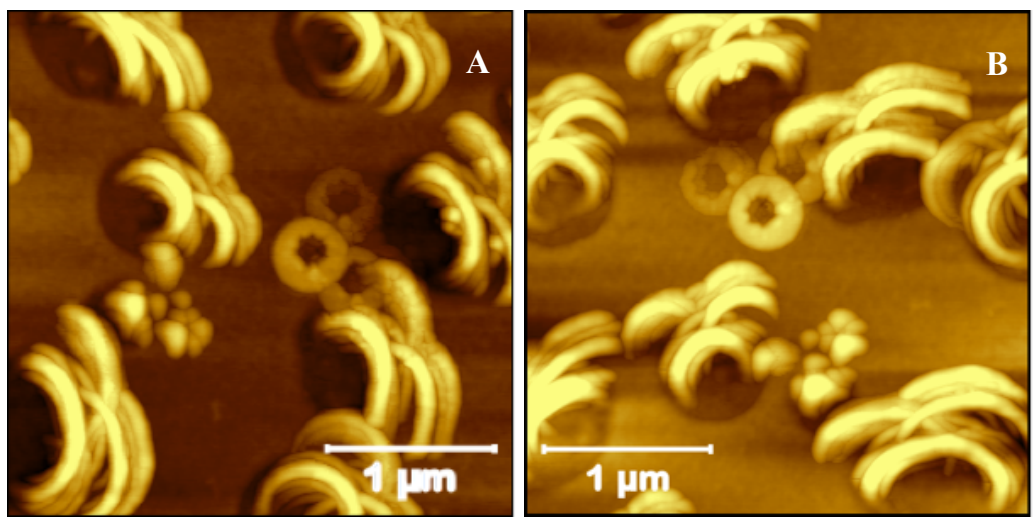

**Figure S18: Effect of scan angle on the gold nanorings**: The same sample region imaged at a scan angle of 0 degree and 90 degrees is shown in Figure A and Figure B, respectively. The rotation of the imaged features in 90 degrees scan as compared to Figure A again eliminates any concerns regarding AFM imaging artifacts.

**Figure S19.**

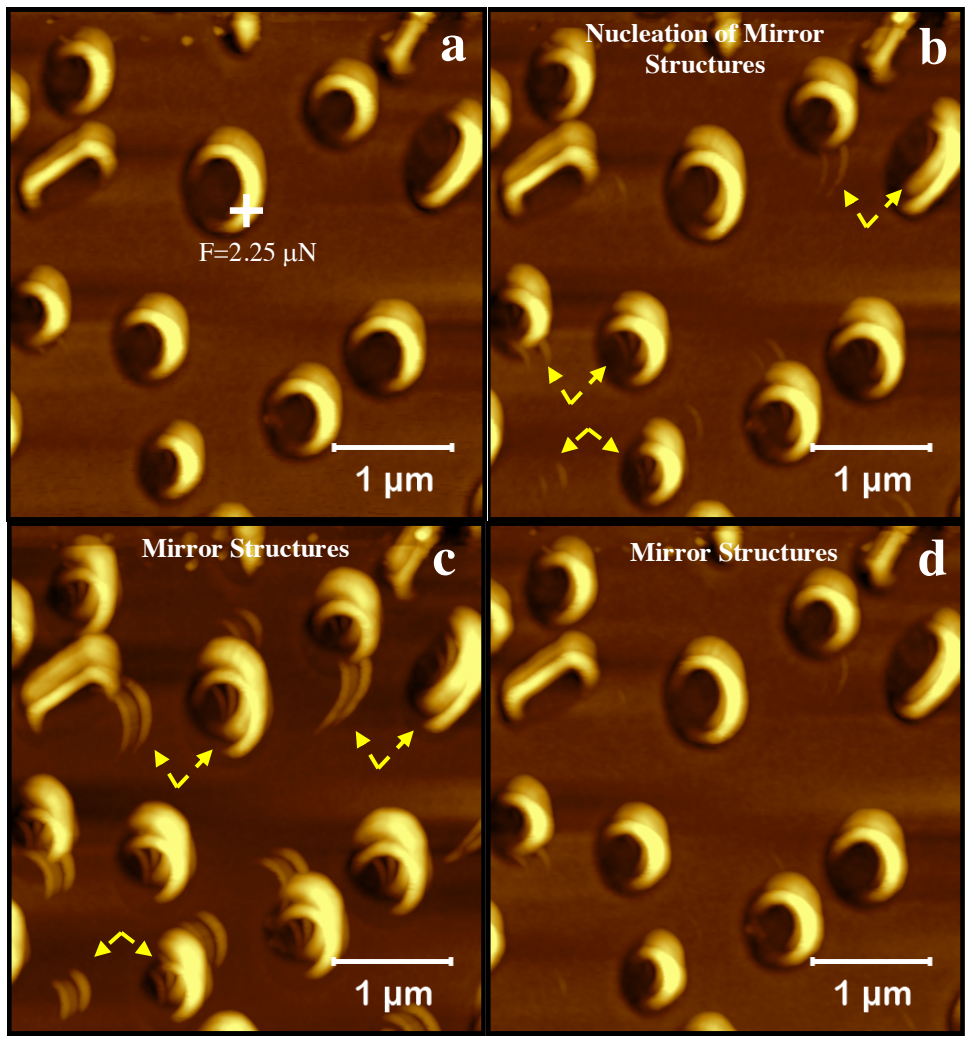

**Figure S19: Observation of the force response *mirroring* behavior in nanorings.** S19a is an AFM image of nanorings prior to the application of a point force of 2.25 µN at the white (+) mark on a nanoring. S19b is the image of the same field immediately after the application of the force. It shows the nucleation of the *mirror* structures, as indicated by the dotted yellow lines with arrows. S19c is an image the same field after ~30 minutes. It shows well-formed *mirror* nanoring structure for every nanoring in the image field. All the mirror structures are aligned

with each other. In S19d we show the image of the same field after approximately an hour and no mirror nanorings were observed.

**Figure S20.**

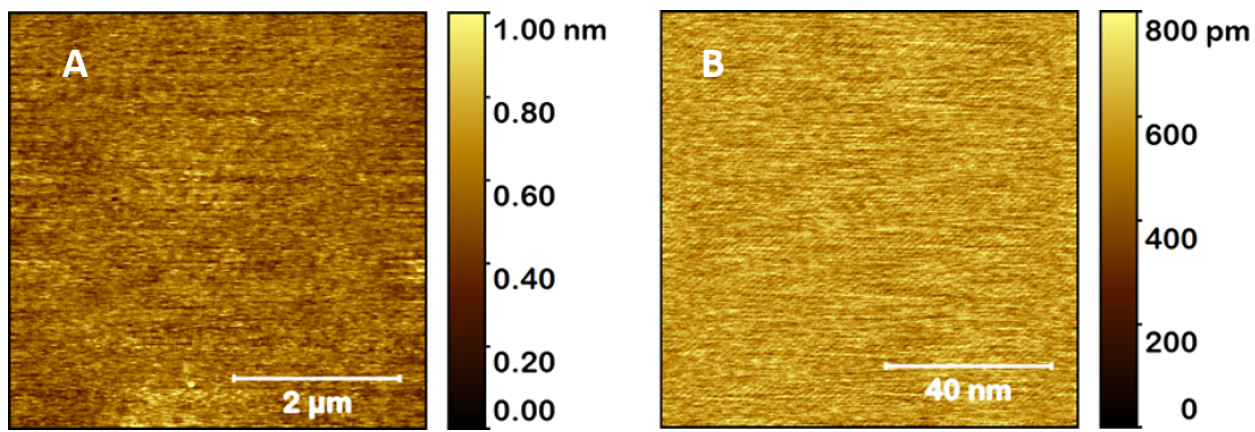

**Figure S20: AFM image of cleaned sapphire substrate:** A is a 5 μm scan and B is 100 nm scan. The average roughness values are found to be as follows: Ra=102 pm, Rrms=130 pm for the 100nm scan shown in B. The sapphire substrate used in our studies was always plasma cleaned prior to gold film deposition.